\documentclass[iop,apj,tighten]{emulateapj}
\usepackage[para,online,flushleft]{threeparttable}
\usepackage{adjustbox}
\usepackage{etoolbox}
\usepackage{tablefootnote, natbib}
\usepackage{aas_macros}
\usepackage{longtable}

\BeforeBeginEnvironment{appendices}{\clearpage}
\usepackage{tabularx}

\newcommand\clearrow{\global\let\rowmac\relax}
\clearrow

\makeatletter
\newcommand\footnoteref[1]{\protected@xdef\@thefnmark{\ref{#1}}\@footnotemark}
\makeatother

\appto\TPTnoteSettings{\footnotesize}
\usepackage[breaklinks,colorlinks,citecolor=cyan,linkcolor=cyan]{hyperref} 

\usepackage[all]{hypcap} 
\usepackage{graphicx}
\usepackage{relsize}

\shorttitle{Luminosity Models and Density Profiles of Nuclear Star Clusters}
\shortauthors{Pechetti et al.}

\begin{document}
\title{Luminosity Models and Density Profiles for Nuclear Star Clusters for a Nearby Volume-Limited Sample of 29 Galaxies}
\author{Renuka Pechetti\altaffilmark{1}, Anil Seth\altaffilmark{1}, Nadine Neumayer\altaffilmark{2}, Iskren Georgiev\altaffilmark{2}, Nikolay Kacharov\altaffilmark{2}, Mark den Brok\altaffilmark{3}}
\affil{\begin{flushleft}
\textit{
{\scriptsize$^1$Department of Physics and Astronomy, University of Utah, 115 South 1400 East, Salt Lake City, UT 84112, USA\\
\scriptsize$^2$Max Planck Instit\"ut f\"ur Astronomie, Heidelberg, Germany\\
\scriptsize$^3$Leibniz-Institut f\"ur Astrophysik Potsdam, An der Sternwarte 16, 14482 Potsdam, Germany\\} 
}\end{flushleft}}

\begin{abstract}
 Nuclear Star Clusters (NSCs) are dense clusters of stars that reside in the centers of a majority of the galaxies. In this paper, we study the density profiles for 29 galaxies in a volume-limited survey within 10 Mpc to characterize their NSCs. These galaxies span a 3$\times$10$^8$ -- 8$\times$10$^{10}$~M$_\odot$ and a wide range of Hubble types.  We use high-resolution Hubble space telescope archival data to create luminosity models for the galaxies using S\'ersic profiles to parameterize the NSCs. We also provide estimates for photometric masses of NSCs and their host galaxies using color--M/L relationships and examine their correlation. We use the Multi-Gaussian Expansion (MGE) to derive the NSC densities and their 3-D mass density profiles. The 3-D density profiles characterize the NSC densities on the scales as small as $\sim$~1~pc, approaching the likely spheres of influence for black holes in these objects. We find that these densities correlate with galaxy mass, with most galaxies' NSCs being denser than typical globular clusters. We parametrize the 3-D NSC density profiles, and their scatter and slope as a function of galaxy stellar mass to enable the construction of realistic nuclear mass profiles. These are useful in predicting the rate of tidal disruption events in galaxies. We will verify the results of this paper in a follow-up paper that presents the dynamical modeling of the same sample of NSCs.  
\end{abstract}

\maketitle

\section{Introduction}
Galactic centers are regions of extreme activity. Some galaxies harbor supermassive black holes (SMBHs) and active galactic nuclei (AGN) while others are undergoing extreme star formation at their centers. Nuclear star clusters (NSCs) are found at the centers of a majority of low-mass and intermediate-mass galaxies ($<$~3$\times$10$^{10}$~M$_\odot$) but are less frequently at the lowest and highest galaxy masses \citep[e.g,][]{seth08,georgievboker14,sanchez19}. These NSCs are the densest known stellar systems with effective radii of $\sim$2-10~pc \citep[e.g,][]{boker04,cote06,georgievboker14} and dynamical masses ranging from 10$^5$ -- 10$^8$~M$_\odot$ \citep[e.g,][]{phillips96,walcher05}. Decomposing NSCs from their host galaxies is essential for characterizing their properties.  Due to their compactness, we can only resolve NSCs in nearby galaxies (D $\lesssim$ 40 Mpc), and using the highest resolution data primarily from the \textit{Hubble Space Telescope (HST)}.

\par 
Many studies have used  $HST$ data to characterize NSCs and understand how their properties scale with galaxy mass and type. Studies of spiral galaxies have revealed that the occupation fraction of these galaxies is high ($\sim$75\%), especially for the latest-type galaxies where it is most straightforward to find NSCs \citep[e.g,][]{carollo02,boker02}. A recent ground-based study in Virgo cluster by \citet{sanchez19} shows that nucleation fraction peaks at roughly 90\% in early-type galaxies with masses $\sim$10$^9$~M$_\odot$ and declines monotonically for both higher and lower mass galaxies. Spectroscopic studies of nearby NSCs show that they contain multiple stellar populations of different ages \citep[e.g,][]{walcher06,rossa06,kacharov18}. Other studies have focused on the masses and effective radii of NSCs and how these correlate with host galaxy properties \citep[e.g,][]{ferrarese06,walcher05,wehner06,turner12,scott13,georgievboker14,georgiev16,savorgnan16}. These NSC scaling relations can add important clues to decipher the fundamental physical mechanisms that drive these relations. For example, they can provide information regarding different formation scenarios of NSCs (such as in-situ formation, cluster mergers \citep[e.g,][]{mclaughlin06,antonini15,gnedin14}. There is a need to accurately determine these relations to improve our understanding of the possible evolution of NSCs along with their host galaxies.

\par Most of the existing NSC mass estimates \citep{seth08,georgiev16,spengler17,sanchez19} are photometric, usually based on SED fitting or color--mass-to-light ratio relations of galaxies. These are not the best indicators of mass due to the age-metallicity degeneracy \citep{worthey94} and because NSCs contain complex star formation histories \citep[e.g,][]{walcher06,siegel07}. Hence, accurate dynamical mass measurements are required to improve the scaling relations of NSCs. One can measure NSC masses by observing the central kinematics of a galaxy. Currently, only a handful of these dynamical measurements exist ($\sim$14) \citep{walcher05,seth10,lyubenova13,denbrok15,nguyen18}. Obtaining additional dynamical masses is important for constraining the scaling relations and quantifying the fraction of galaxy mass that forms the dense NSC. 

\par  NSCs often co-exist with SMBHs, especially in galaxies with masses M$_{gal}$$\sim$10$^{10}$M$_\odot$ \citep{filippenkoho03,seth08,graham09,seth10,neumayerwalcher12,nguyen18}, but the scaling of their masses with galaxy properties appear to be distinct \citep[e.g.][]{erwingadotti12,scott13}. A consequence of co-existence of NSCs with SMBHs is the occurrence of luminous Tidal Disruption Events (TDEs). When a star from the NSC comes within the Roche radius of the BH and roughly half of its mass falls onto the BH generating a tidal disruption flare \citep[e.g,][]{hills75,rees88,vanvelzen11,arcavi14,tadhunter17}. Recent studies like \citet{graur18,dorazio19} have shown that the rate of TDEs depends on the stellar density. The TDE rates of galaxies change based on the number of stars surrounding the BH. It can vary either quadratically or linearly based on the emptiness or fullness of the loss-cone regime \citep[e.g.,][]{stone16}. The orbits of stars that matter most in estimating these rates are those within the sphere of influence (SOI) radius of the black hole. Thus, it is necessary to measure the NSC density at high spatial resolution on the scale of the SOI to probe the TDE rate measurements in theoretical work \citep[e.g][]{ivanov05,chen09,stone16}, as well as matching the density measurements as closely as possible to the mass distribution of galaxies in the local universe.

\par Measuring the densities of NSCs can be a key to constrain the theoretical TDE rates more accurately in galaxies, which in turn helps us in understanding the role of tidal disruptions in the evolution of NSCs along with their galaxies. Currently, it is challenging to determine observational TDE rates as they are challenging to detect ($\sim$~40 observed)\citep[e.g,][]{arcavi14,french16,wevers19}. TDEs are usually detected in real-time transient surveys like All-Sky Automated Sky Survey (ASASSN), Palomar Transient Factory (PTF), etc.  In future, we would be able to observe these with surveys like Large Scale Synoptic Telescope (LSST). This enables us to predict the TDE rates and compare them with theoretical predictions \citep{stone16,vanvelzen18}. In addition to TDEs, the density profiles of NSCs may be important for predicting binary merger sources of gravitational waves \citep[e.g.][]{fragione19,antonini19}.

Motivated by the need to measure the innermost density profiles and total dynamical masses of NSCs, we analyzed a sample of 29 galaxies with high-resolution data in the $\it{HST}$ archive. These were observed with GEMINI/GNIRS and VLT/XHOOTER to obtain their central kinematics.

In this paper, we characterize the properties of these galaxies and their central light and mass profiles estimated from stellar colors.  In a follow-up paper, we will present the dispersion measurements of each NSC and their dynamical masses derived using the luminosity models presented in this paper. Section~\ref{sec:data} describes the sample selection, $HST$ data, and our galaxy masses. In Section~\ref{sec:hrmodels}, we introduce our method to derive the luminosity models for all galaxies and present their Multi-Gaussian Expansion (MGE) models. In Section~\ref{sec:color_masses} we describe the method to derive the NSC masses and present the correlations between them and their host galaxies. Section~\ref{sec:densities} shows the central densities and 3-D density profiles of NSCs and discusses their connection to TDEs. We conclude in Section~\ref{sec:conclusion}.

\begin{figure}[!ht]
\centering
\includegraphics[width=\linewidth]{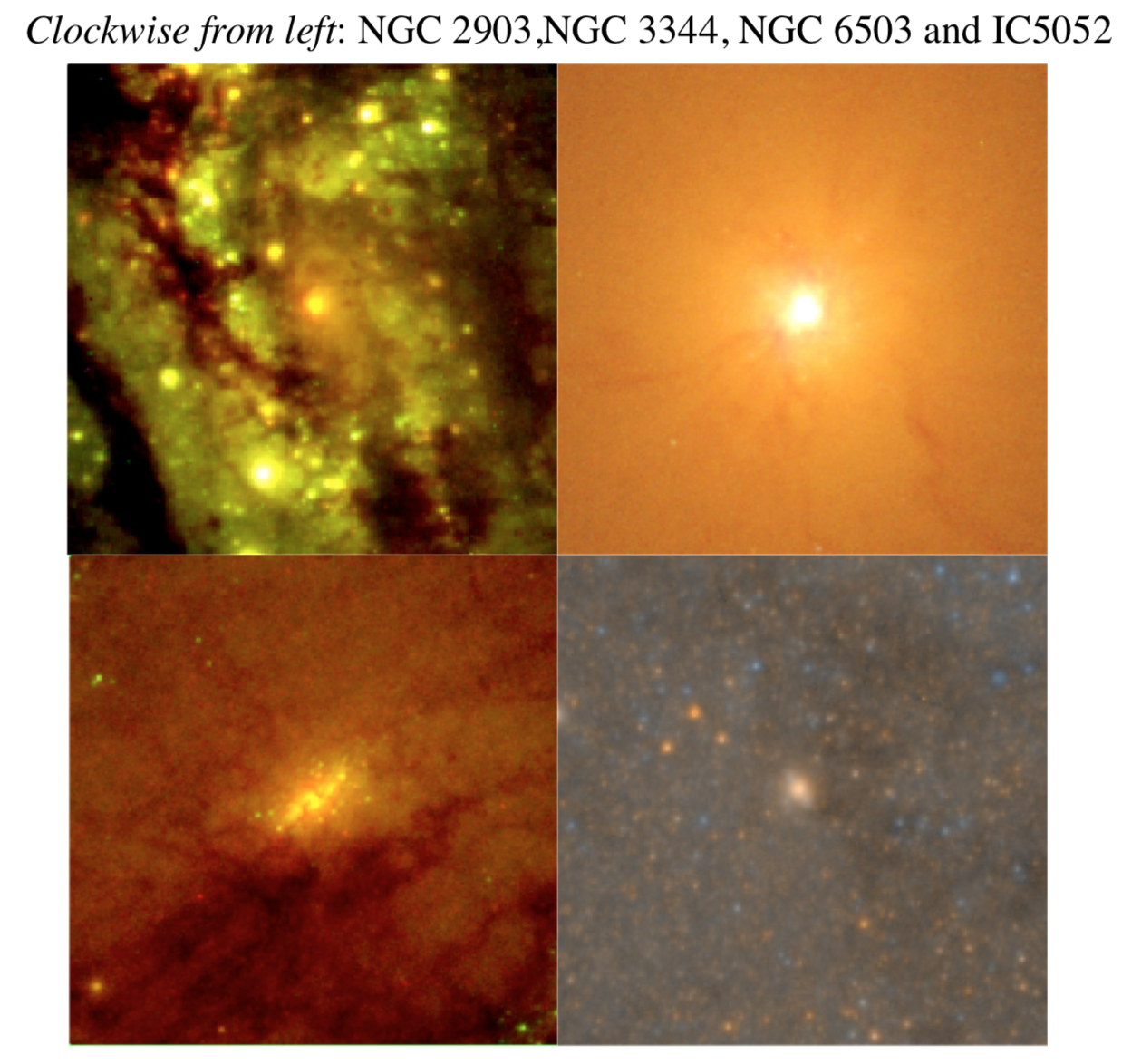}
\caption {A representative sample of our galaxies from $HST$ is shown here. The images show the 10$''$~$\times$~$10''$ central region of the galaxies with pixel scales ranging from 0.04$''$ - 0.05$''$~pixel$^{-1}$. Clockwise from top left: NGC~2903 (WFPC2/PC), NGC~3344 (WFC3/UVIS), NGC~6503 (WFC3/UVIS) and IC~5052 (ACS/WFC).}
\label{fig:galaxies}

\end{figure}
\begin{figure*}[!ht]

\epsscale{1.2}
\includegraphics[width=\linewidth]{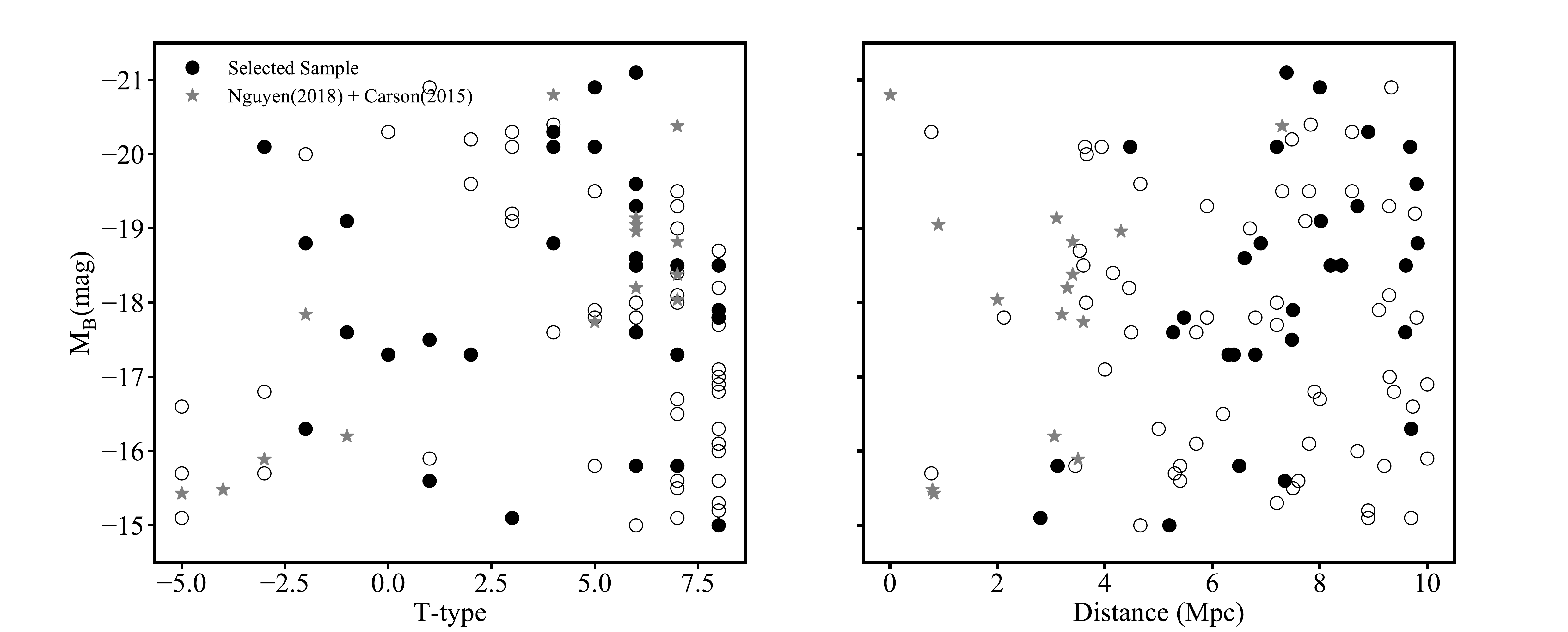}

\caption{Our sample of 29 galaxies with NSCs (solid points) is drawn from the sample of galaxies within 10 Mpc and with $M_B < -15$ \citep{tully88,karachentsev04} (all points).  Gray stars indicate the  15 nearby galaxies  with NSC properties derived in \citet{carson15,nguyen18}  {\em Left:} The luminosity vs. Hubble type of the sample, {\em Right:} The luminosity vs. distance.}
\label{fig:mag_t}
\end{figure*}

\section{Sample Selection and Data}
\label{sec:data}
In this paper, we create luminosity models with high-resolution $HST$ imaging to quantify the morphology and properties of NSCs in these galaxies. Here we discuss the sample selection and the $HST$ data available for our sample. We also describe the estimation of galaxy stellar masses. 

\subsection{Sample Selection}

\par We restricted our sample to galaxies within 10 Mpc for several reasons. First, for a galaxy at 10~Mpc, a typical NSC with an effective radius of $\sim$5~pc corresponds to 0.1$''$. Given the resolution of $HST$ (FWHM~0.07$''$), 10~Mpc, therefore represents the outer edge where we can resolve NSCs. We also want to resolve the SOI (= G.M$_{BH}$/$\sigma^2$) of black holes with M$_{BH}$$\sim$10$^6$~M$_\odot$, which also corresponds to $\sim 0 \farcs 1$ at 10~Mpc for a galaxy with $\sigma$~$\sim$~30~km/s. This is important for any future dynamical BH mass measurements from high-resolution stellar or gas dynamics using telescopes like ALMA and JWST. Second, it becomes challenging to separate the galaxy from the NSC in ground-based spectroscopy, and thus difficult to measure the dynamical masses of clusters beyond this distance \citep{walcher05}. 
Third, we would like to quantify the density of stars at radii relevant for predicting TDE rates, so that these densities can be used to improve TDE rate estimates in more distant galaxies. Our goal is to get precise dynamical masses and densities, and thus 10~Mpc makes a good conservative distance limit. 

\par We constructed the sample considering all galaxies of type from E through Sd with a B-band absolute magnitude of less than -15 from the \citet{karachentsev04} and \citep{tullyfisher88} catalogs of nearby galaxies. 
The sample was further restricted to galaxies with $HST$ data. We removed well-studied NSCs in nearby ($<$~5~Mpc) early-type galaxies from the sample of \citet{nguyen18,nguyen19} and late-type galaxies sample from \citet{carson15}. The results from these studies will be in some of our analyses below.  Finally, we examined the remaining galaxies for distinct NSCs using both $HST$ and 2MASS data; the selected NSCs have $K < 15$ mag. We focus on the properties of NSCs here and plan to focus on the occupation fraction of NSCs in a forthcoming paper (Hoyer et al., in prep). $HST$ color images of a few representative examples for the complexity in their central regions are shown in Figure~\ref{fig:galaxies}. They are 10$''$$\times$10$''$ color images from different cameras of $HST$ and have pixel scales ranging from 0.04$''$ - 0.05$''$~pixel$^{-1}$. Figure~\ref{fig:mag_t} shows the parent sample and our selected sample with respect to their T-type (Hubble-type) and distance, which consists of 29 galaxies. Galaxies from \citet{carson15,nguyen18,nguyen19} are also shown. The majority of our sample consists of late-type galaxies, while 5 out of 29 are early-type galaxies. 

\par To derive dynamical masses of NSCs in these galaxies we need spectroscopic data at their centers. We acquired long-slit spectroscopic data for the 29 galaxies from GEMINI/GNIRS and VLT/XSHOOTER. We describe the data reduction and the analysis in the second series of this paper, where we derive the integrated velocity dispersions for the central NSC and determine their dynamical masses and densities using the luminosity models presented in this work.

\subsection{$HST$ data}
\par We used high-resolution $HST$ data for modeling the NSCs in these galaxies as they are very compact (effective radii$\sim$2-10 pc) that corresponds to $\sim$0.06-0.25$''$ at the median distance ($\sim$8 Mpc) of our sample. The
$HST$ pixel scale is smaller than the effective radii of the NSCs. We used high-resolution imaging from WFPC2, ACS/WFC, WFC3/UVIS and WFC3/IR cameras. We downloaded the individual \texttt{.flt/.flc} files (ACS and WFC3 data) and \texttt{.c0m} files (WFPC2 data) from the $Hubble$ archive and drizzled them using \texttt{Astrodrizzle} \citep{drizzlepac} to the desired pixel scales. We used the following pixel scales for different cameras. For WFPC2/PC -- 0.045$''$~pixel$^{-1}$, WFC3/UVIS -- 0.04$''$~pixel$^{-1}$, ACS/WFC -- 0.05$''$pixel$^{-1}$, WFC3/IR -- 0.08$''$~pixel$^{-1}$ and WFPC2/WF -- 0.1$''$~pixel$^{-1}$. 

\par We restricted our analysis to data from filters~$\geqslant$~5346\AA~ to reduce the uncertainties due to stellar population variations and dust. Table~\ref{table:HST} lists the cameras and filters we used to model the data. We used the F814W filter wherever possible (18 out of 29 galaxies), as this filter retains the higher resolution of the optical cameras but is the reddest filter typically available. Dust was masked using several techniques as described in Section~\ref{sec:dust}. We did not use the data that was saturated or contained bad pixels at the very center as that would create inaccurate models. In the case of very dusty galaxies, where dust masks were not useful, we used WFC3/IR F160W/F110W observations (for NGC~5194, NGC~5195, NGC~5236), as it features much less dust compared to the F814W band and is, therefore, easier to model. We note, however, that the NSC structure could differ with increasing wavelength, as studies have found evidence for that \citep[e.g][]{georgievboker14,carson15}. Table~\ref{table:params} in the  lists the sample of galaxies along with some useful quantities. We also list some galaxies in the Appendix that required additional attention while modeling their NSCs.

\subsection{Deriving galaxy masses}
\label{sec:galmass}
\par An important goal of this study is to analyze how the structure of NSCs varies with the stellar mass of their host galaxies. A majority of our galaxies (18 of 29) had existing stellar mass estimates. These were taken from \citet{cook14}, where they used $Spitzer$ 3.6 $\mu$m luminosities and a mass-to-light ratio ($M/L$) of 0.5 at 3.6 $\mu$m to estimate the galaxy masses. A variety of methods such as estimating masses using star formation histories \citep[e.g,][]{Oh08,eskew12,mcgaugh14,barnes14,meidt14} have consistently predicted a $M/L_{3.6\mu}$$\sim$0.5. Hence, we used these mass estimates for 18 galaxies. 

\par For the remaining galaxies (11 of 29), we used the following data to estimate galaxy stellar masses using their integrated colors and magnitudes: 
\begin{itemize}
\item For total galaxy luminosities, we used K-band magnitudes from the 2MASS Large Galaxy Atlas \citep{jarrett03}. 
\item The distances were obtained from the \citet{karachentsev04} updated catalog of nearby galaxies\footnote{We note that the galaxy NGC~4941 is at a distance of $\sim$15~Mpc. Its distance estimate was within 10~Mpc when the sample was selected initially. The \citet{karachentsev04} updated catalog later removed the galaxy. We still present the analysis for this galaxy even though its outside our distance limit.}. We used the foreground extinction A$_v$ from Nasa Extragalactic Database (NED), which is based on the \citet{schlafly11} extinction map.
\item For four out of eleven galaxies, we used $g-i$ colors, which were available in the NASA Sloan Atlas\footnote{\tt http://www.nsatlas.org/data}. \item For the remaining seven non-SDSS galaxies, we used $B-V$ colors from HYPERLEDA\footnote{\tt http://leda.univ-lyon1.fr/} that were derived within an effective aperture radius. 
\end{itemize}
From these colors, we estimated $M/L$s of the galaxies using the relations described in \citet{roediger15} and derived their corresponding masses. These color-M/L relations were derived from stellar population synthesis (SPS) models and using multi-band photometry. \texttt{MAGPHYS} library was used to compute the stellar SEDs based on the \citet{bc03} (BC03) and \citet{fsps} (FSPS)  SPS models, which assume a Chabrier IMF (see \citet{roediger15} for more details on the derivation of these relations). We used the color-$M/L$ relations based on the BC03 models. The galaxy masses that we derived are listed in the Appendix Table~\ref{table:params}. All of the colors and magnitudes are expressed in the AB magnitude system.

\par To estimate the errors on our galaxy masses we compared the masses measured from $B-V$ and $g-i$ colors. We used the galaxies in our full sample for which both the colors were available (13 galaxies) and estimated their masses. We found them to vary within $\sim$25\%.  Another source of error could arise from the choice of colors that we used to measure the masses. To estimate that, we calculated the masses using $g-r$ colors using similar relations from \citet{roediger15} and then computed the difference with the previously estimated masses that used $g-i$ colors. Our errors were of the order of $\sim$~6\%, which were much smaller than the earlier error estimate.  Distance errors add to the errors in our mass estimates. We obtain a median fractional error on all the distances in our sample, which is $\sim$5.7\% and incorporate this median error. Also, we propagate the error from the color-M/L relation ($\sim$10\% from \citet{roediger15}) to estimate the total error on the masses. We find a median error of 15.5\%, and we add this error in quadrature to the previous error estimate of 25\% to get the final error estimate of $\sim$30\% on our masses.  
For the 18 galaxy mass estimates available from \citet{cook14} we find consistent estimates within 1$\sigma$ for 80\% of these galaxies, while the remaining were within 2$\sigma$.

\begin{figure}[ht]

\centering
\includegraphics[width=0.54\textwidth]{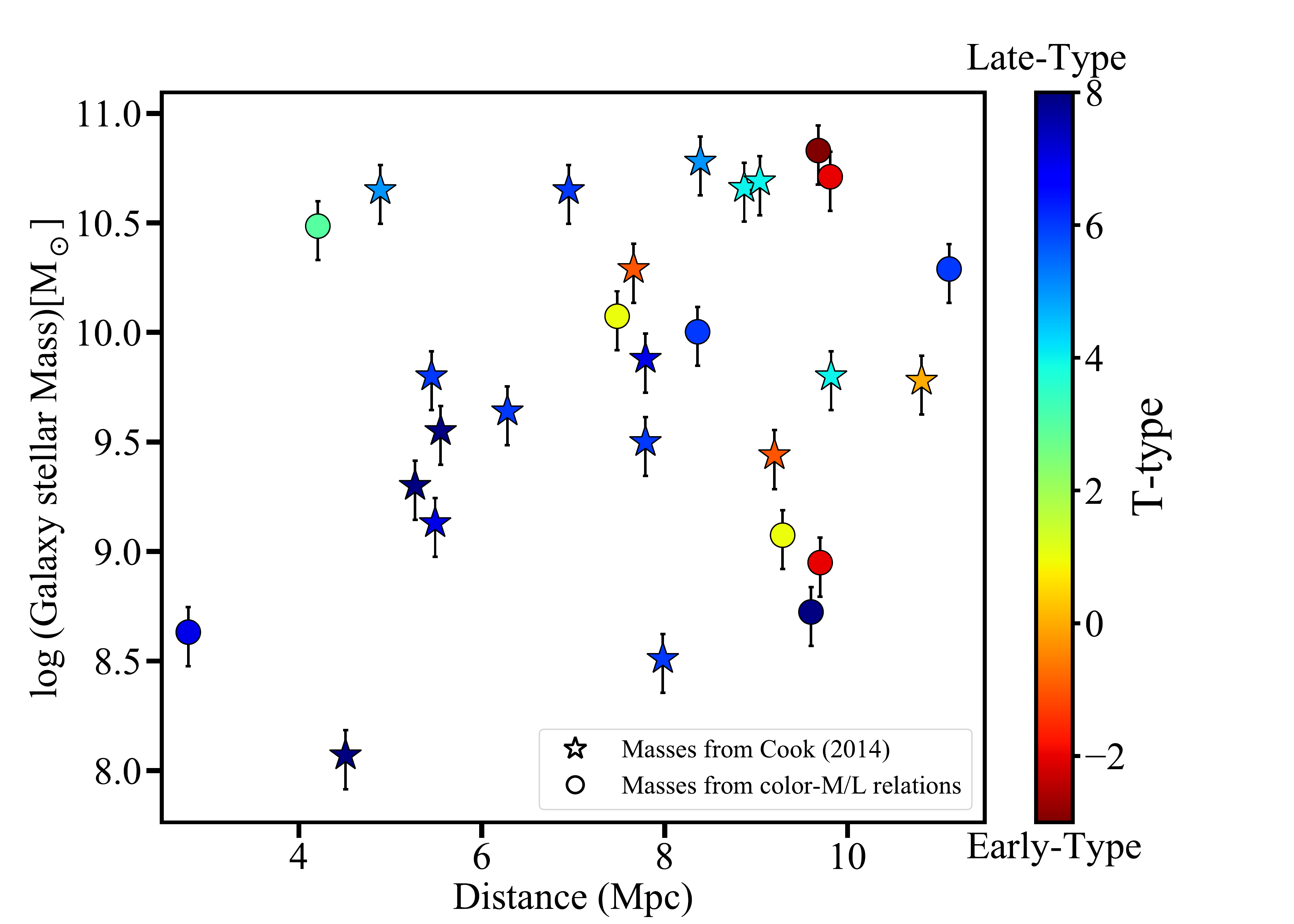}

\caption {Thew galaxy stellar masses and distances of our galaxy sample.  Stellar masses are derived from \citet{cook14} or color-M/L relations from \citet{roediger15}. The galaxies are colored with respect to their T-type (Hubble-type). The reddest points are early-type galaxies and the bluer points are late-type galaxies.}
\label{fig:mass_dist}

\end{figure}

Figure~\ref{fig:mass_dist} shows the masses for our galaxies with respect to their distances and colored according to their Hubble-type. The majority of our sample are late-type galaxies which are mostly in blue, and the early-types are in red.  The stars are the masses derived from \citet{cook14} and the circles are our new galaxy mass estimates. 

\section{Creating high-resolution luminosity models}
\label{sec:hrmodels}
The primary goal of this paper is to measure the mass profiles of NSCs on scales of $\geq$~1~pc. In this section, we first discuss how we determine the $HST$ Point Spread Function (PSF) for all the images, which will be used in convolving models to match our images. Then, we describe the derivation of luminosity models and the determination of their errors. We also discuss different masking techniques employed in the estimation of these models. 

\subsection{PSF determination}
 
We require an accurate PSF model as our focus is on the central regions of galaxies, where the PSF effects are significant. We used the \texttt{Tiny Tim} code suite \citep{kristhook11} to generate PSFs for our galaxies. The \texttt{tiny1} code was used to create a parameter file using the input parameters such as the camera, filter, size, etc. for the PSF. The \texttt{tiny2} code was used to create an undistorted PSF using the parameter file. To account for the spatial variability of PSFs in an $HST$ frame (ACS and WFC3 data), we used the individual frames (\texttt{.flt/.flc} files) of a galaxy to get the positions of galaxy centers. Then, we created PSFs in the same position using the \texttt{tiny3} code. Later, we used \texttt{Astrodrizzle} \citep{drizzlepac} and drizzled the individual dithered PSF images to a final PSF using the same drizzle parameters and pixel scale as the galaxy image. For WFPC2 data, we used the undistorted PSF as there was no significant correction required using the \texttt{tiny3} code. These PSFs were then later used for fitting the galaxy light profile to deconvolve them from the galaxy. 

\subsection{Surface photometry of the galaxies}
We fit the luminosity of 29 galaxies in 2-D and created their models.  We first describe the creation of the masks used to remove stars, dusty areas, and bad pixels from the fits.  We then describe the best-fit models on the surface brightness and the errors on these best-fit models estimated using bootstrapping.

\subsubsection{Creating masks for modeling}
\label{sec:dust}
\par In some of the galaxies, dust and foreground stars hinder our ability to make an accurate luminosity model. If there was any visible dust in the images, we created dust masks using color maps of the galaxy. For creating the color maps, we used data in two filters for each galaxy, typically with a redder filter like F814W and a bluer filter like F555W. Then we created a mask by masking all the regions in the colormap above a threshold level. This works well if the light coming from behind the dust is still the dominant light in the image. However, in some cases, the dust extinction was so high that little background light penetrated, and thus the colors of these regions reflected the foreground stellar populations.  
In these cases, the dust lanes were not easily identified using color maps, so we took an alternative approach that was also used in \citep{georgiev19}. We used the IRAF task ELLIPSE \citep{jedrzejewski87}, which fits elliptical isophotes to galaxy images using parameters such as position angle, ellipticity, and intensity for each isophote. We then created a model image using the task BMODEL that uses the output from the ELLIPSE task. We then subtracted the original science image from the model ELLIPSE image to create a residual image, where we used some threshold level to create a mask for the galaxies. 

Foreground and bright stars that contaminated galaxies were masked too as they could dominate the galaxy's light, and hence overestimating the light in the models. We did not mask the central regions ($\sim$0.4$''$) in the images even if there was visible dust/extinction, as it was crucial for modeling the NSC.

\subsubsection{Fitting the Surface Brightness Profiles}
\label{sec:imfit}
\par The NSC in a galaxy is usually identified by the visible bump in the surface brightness profile \citep[e.g.][]{cote06} on scales of $\lesssim$~50~pc. To quantify the NSC parameters, we need to decompose the light of the galaxy into different components. We used S\'ersic profiles to deconstruct the surface brightness profiles of the NSCs in our galaxies. This choice was made to maintain homogeneity across our sample and also to facilitate comparisons to existing measurements NSC S\'ersic profiles \citep{graham09,carson15,ngyuen17,nguyen18}. The first generalized S\'ersic $r^{1/n}$ model was proposed by \citet{sersic68}. The standard S\'ersic model \citep{graham05} is described by 
\begin{equation}
I(a) = I_e~exp\left\{-b_n\left[\left(\frac{a}{r_e}\right)^{1/n}-1\right]\right\}
\end{equation}
where I$_e$ is the surface brightness at the effective (half-light) radius r$_e$ and n is the S\'ersic index, which controls the shape of the intensity profile. b$_n$ is given by the solution to the transcendental equation
\begin{equation}
\Gamma(2n) =  2\gamma(2n,b_n)
\end{equation}
where $\Gamma$(a) is the gamma function, and $\gamma$(a,x) is the incomplete gamma function. Usually, the nuclei are superimposed on the galaxy light and, therefore, are required to be fitted with a 2 component profile fit. Occasionally, we fit a 3 component profile where the galaxy might have a third component of an underlying bulge and disk within the radius that we fit as in \citet{georgiev19}. 

We fit our galaxies using the \texttt{IMFIT} code, as described in \citet{erwin15}. For a S\'ersic profile, the code builds a model 2-D image using the input parameters; intensity (I$_e$), effective radius (R$_e$), S\'ersic index (n), position angle (PA) and the ellipticity ($\epsilon$) and then convolves it with the given PSF. It then modifies the model image parameters and uses non-linear minimization of the total $\chi^2$ to find the closest model of the galaxy. 

We fit a region of 8$''$-10$''$ which corresponds to a region around 300 - 400 pc at a median distance of 8 Mpc. For some galaxies, we fit a two S\'ersic profile fit, whereas some fit better with a S\'ersic NSC and an exponential disk profile. We also fit a component for the flat background wherever necessary, which can accommodate any large uniform component in our galaxy or data. We mostly used F814W data for modeling the galaxies except for a few that either did not have the useful data in F814W, or they were better modeled in IR filters. 

\begin{figure}[ht]

\label{fig:surf_bri}
\includegraphics[width=0.5\textwidth]{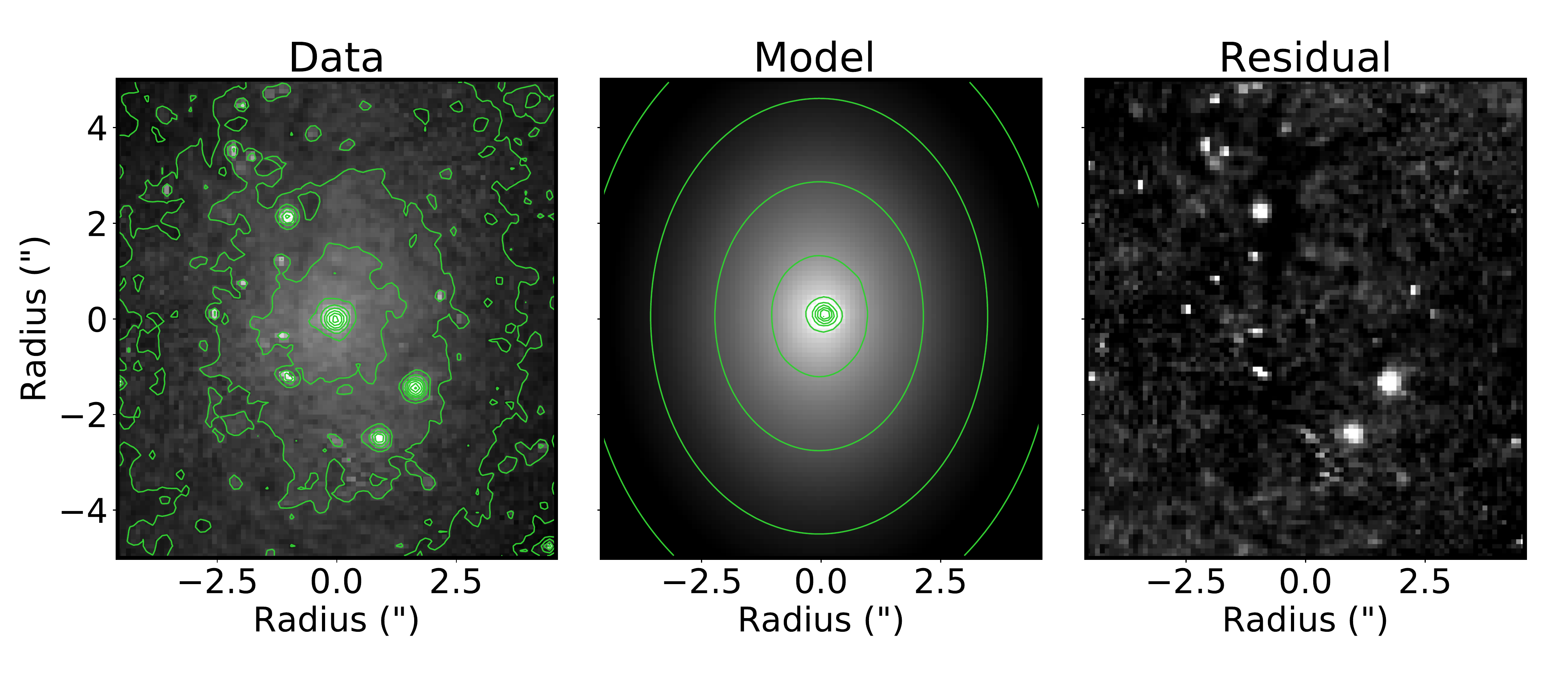}
\includegraphics[width=0.5\textwidth]{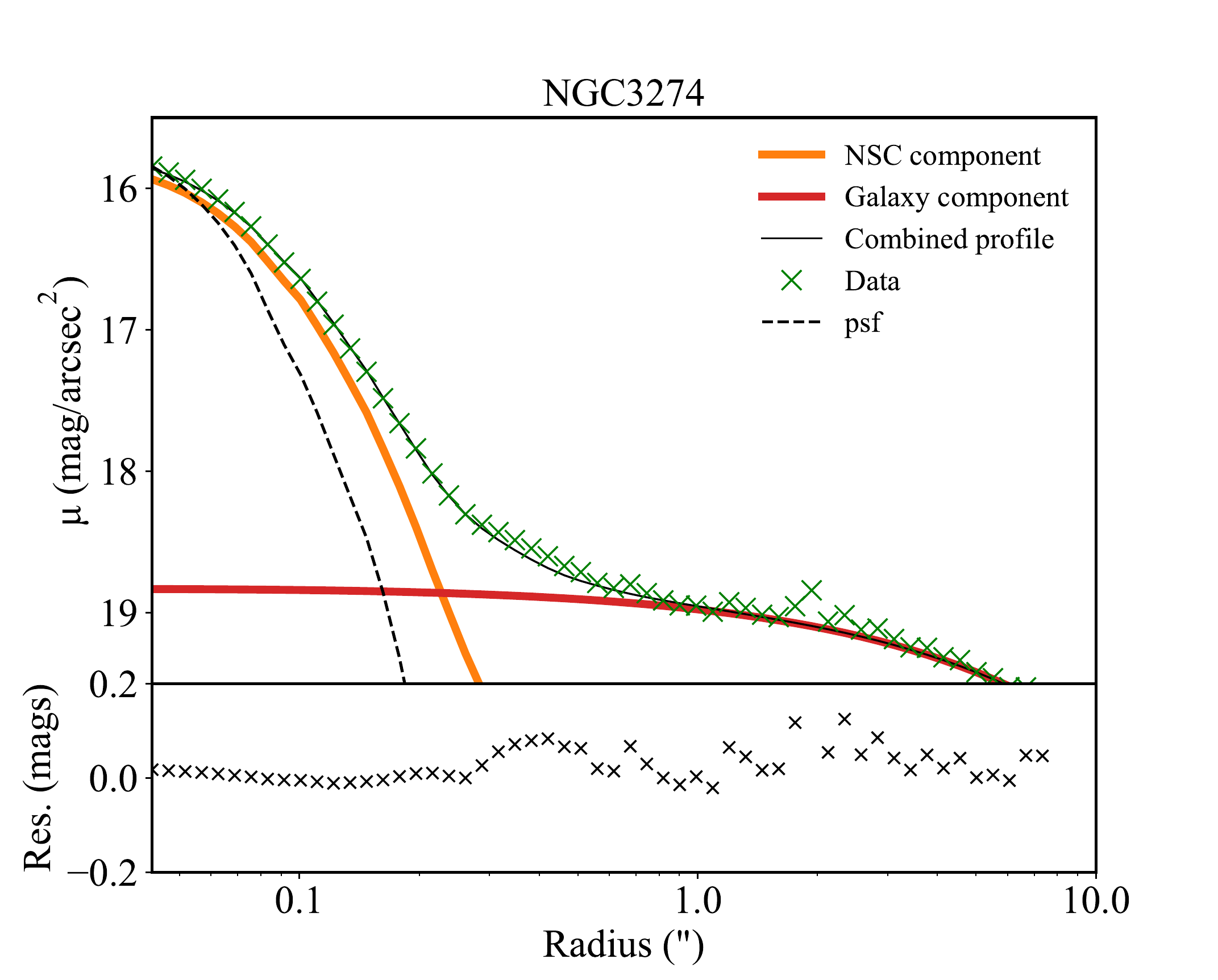}
\caption {\textit{Top Panel}: Data, \texttt{IMFIT} model, and residual image of NGC~3274. A region of 10$''$ is shown here. \textit{Bottom Panel}: 1-D surface brightness profile (in black) of the same galaxy described by two individual S\'ersic profiles (in orange and red). Green points show the data. The upward bump in the surface brightness profile within 1$''$ indicates the presence of an NSC in the galaxy. The black points in the lowest panel show the residual (model-data) in magnitudes. }
\end{figure}

We briefly discuss some of the complications that arose in our fits here. 
\begin{itemize}
\item AGN activity: There were a few galaxies in our sample which are classified as Seyferts, and thus may contain a point-like AGN component. However, we didn't find any galaxies in our final sample where our fit was improved with the inclusion of a point source. Three galaxies are classified as Seyfert IIs (Circinus galaxy, NGC~4941 and NGC~5194).  Despite this, we found no evidence for AGN components in these galaxies, consistent with the lack of detectable AGN variability in most lower luminosity AGN like those we observe here \citep{baldassare18}. 

\item Dust extinction: NGC~3593 has a very red color and affected by a lot of reddening. For three galaxies (NGC~5194, NGC~5195, and NGC~5236), we used F110W, F110W and F160W data respectively for modeling, as the dust obscured a lot of galaxy light in F814W band. In this case, we used the color map derived from F814W and F555W images to create a mask to model the IR data.  Circinus galaxy is also obscured by a lot of dust both internal and foreground due to being close to the galactic plane.
\item High S\'ersic index: Some of our galaxies also have a high S\'ersic index (n$>$6). \citet{carson15} also estimate high S\'ersic indices for some galaxies in their sample, which is due to the behavior of the profile in the wings rather than in the core (see Section~\ref{sec:sersic}. We describe the outlier galaxies in the Appendix. 

\end{itemize}
We assigned a quality to our fits, and even though the dust was masked in all the galaxies, some were not fit well due to the presence of dust close to the nucleus of the galaxy. Quality 0 was assigned for those that had an overall good fit, and the 1-D fit residuals were within 0.1 mags. Quality 1 was assigned for the galaxies that had a good fit closer to the center (NSC) of the galaxy but residuals greater than 0.1 mags further from the central arcsecond. Quality 2 was given to those that had a lot of dust contamination, and the NSC was not fit properly and contained residuals greater than 0.2 mags. Table~\ref{table:sersic} provides the S\'ersic fits for our sample of galaxies along with their quality of fits. Figure~\ref{fig:surf_bri} shows an example of the \texttt{IMFIT} model of NGC~3274. Here, we use a two-component S\'ersic model to fit our galaxy and NSC, with some initial guesses inputs for n, I$_e$, R$_e$, PA, and $\epsilon$ for both the components. We also fit for the flat background, which is subtracted in Figure~\ref{fig:surf_bri}. 

\begin{table*}
\centering
\caption{S\'ersic parameters for the NSCs}
\label{table:sersic}
\begin{threeparttable}
\bgroup
\def\arraystretch{1.8}
\setlength\tabcolsep{5pt}
\begin{tabular}{cccccccccp{0.1\linewidth}}
\hline\hline
Galaxy    & n    &  logI$_e$ (L$_\odot$/pc$^2$)  & R$_e$ (pc)  & PA$_{nuc}$ &  $\epsilon_{nuc}$ & Quality	&PA$_{gal}$ & $\epsilon_{gal}$	& AGN/Dust/Comments\\
(1)  & (2)  & (3)  & (4)  &  (5) & (6)  & (7)  & (8) & (9) \\
\hline
Circinus & 1.09$_{-0.10}^{+0.10}$   & 3.93 $\pm$ 0.02   & 8.00 $_{-0.20}^{+0.27}$   & 160.81$\pm$ 5.68       & 0.28 $\pm$ 0.05 & 1  & --      & --   &SeyferII, extincted    \\
ESO274-1 & 2.28$_{-0.10}^{+0.26}$   & 4.20 $\pm$ 0.04   & 2.18 $_{-0.10}^{+0.10}$   & 77.25 $\pm$ 6.77       & 0.12 $\pm$ 0.02 & 1  & --      & --       \\
IC5052   & 1.77$_{-0.09}^{+0.09}$   & 3.62 $\pm$ 0.01   & 3.73 $_{-0.08}^{+0.08}$   & 149.25$\pm$ 1.41       & 0.25 $\pm$ 0.02 & 0  & 142.7   & 0.783    \\
IC5332   & 7.53$_{-0.99}^{+1.29}$   & 1.88 $\pm$ 0.06   & 23.18$_{-2.31}^{+2.31}$   & 124.37$\pm$ 3.18       & 0.30 $\pm$ 0.06 & 2  & 0.7     & 0.035 & High S\'ersic Index    \\
NGC2784  & 3.98$_{-0.68}^{+0.45}$   & 3.28 $\pm$ 0.08   & 20.37$_{-1.26}^{+2.16}$   & 2.69  $\pm$ 3.06       & 0.61 $\pm$ 0.05 & 1  & --      & --       \\
NGC2787  & 1.02$_{-0.03}^{+0.32}$   & 4.40 $\pm$ 0.06   & 5.12 $_{-0.08}^{+0.08}$   & 54.83 $\pm$ 29.52      & 0.02 $\pm$ 0.03 & 0  & 110.9   & 0.459    \\
NGC2903  & 1.56$_{-0.10}^{+0.10}$   & 3.96 $\pm$ 0.02   & 10.32$_{-0.44}^{+0.44}$   & 22.58 $\pm$ 3.48       & 0.18 $\pm$ 0.02 & 0  & 19.4    & 0.508    \\
NGC3115  & 4.51$_{-0.41}^{+0.19}$   & 3.92 $\pm$ 0.09   & 26.89$_{-2.68}^{+2.68}$   & 44.95 $\pm$ 0.45       & 0.54 $\pm$ 0.03 & 0  & 38.7    & 0.38     \\
NGC3115B & 1.40$_{-0.08}^{+0.08}$   & 3.94 $\pm$ 0.01   & 6.61 $_{-0.09}^{+0.09}$   & 19.70 $\pm$ 1.09       & 0.19 $\pm$ 0.01 & 0  & --      & --       \\
NGC3184  & 5.22$_{-1.42}^{+1.62}$   & 3.80 $\pm$ 0.44   & 2.05 $_{-0.26}^{+0.26}$   & 105.45$\pm$ 6367257.12 & 0.00 $\pm$ 0.29 & 1  & --      & 0.145    \\
NGC3274  & 4.99$_{-0.13}^{+0.41}$   & 3.58 $\pm$ 0.05   & 3.32 $_{-0.22}^{+0.22}$   & 85.75  $\pm$ 2.95       & 0.24 $\pm$ 0.02 & 0  & 96.1    & 0.464    \\
NGC3344  & 0.78$_{-0.06}^{+0.06}$   & 5.03 $\pm$ 0.01   & 4.79 $_{-0.11}^{+0.11}$   & 10.64 $\pm$ 4.40       & 0.16 $\pm$ 0.02 & 0  & 9.74    & 0.1      \\
NGC3593  & 1.40$_{-0.14}^{+0.14}$   & 4.84 $\pm$ 0.10   & 5.50 $_{-0.23}^{+0.23}$   & 95.22 $\pm$ 7.92       & 0.30 $\pm$ 0.05 & 2  & 87.6    & 0.499 & highly extincted   \\
NGC4242  & 6.77$_{-0.35}^{+0.50}$   & 3.93 $\pm$ 0.03   & 1.74 $_{-0.06}^{+0.06}$   & 175.82$\pm$ 5.43       & 0.08 $\pm$ 0.01 & 0  & 156.2   & 0.286   &High S\'ersic Index  \\
NGC4460  & 0.92$_{-0.08}^{+0.08}$   & 4.45 $\pm$ 0.03   & 4.67 $_{-0.10}^{+0.10}$   & 67.3  $\pm$ 1.33       & 0.47 $\pm$ 0.02 & 0  & --      & 0.639    \\
NGC4517  & 4.90$_{-0.43}^{+0.43}$   & 3.63 $\pm$ 0.04   & 5.73 $_{-1.38}^{+1.38}$   & 91.18 $\pm$ 1.74       & 0.46 $\pm$ 0.03 & 1  & --      & --       \\
NGC4592  & 9.88$_{-0.78}^{+0.78}$   & 3.90 $\pm$ 0.09   & 3.45 $_{-0.33}^{+0.33}$   & 61.97 $\pm$ 3.40       & 0.27 $\pm$ 0.04 & 1  & 94.1    & 0.691 &High S\'ersic Index    \\
NGC4600  & 2.10$_{-0.20}^{+0.49}$   & 3.76 $\pm$ 0.05   & 8.27 $_{-0.73}^{+0.73}$   & 37.09 $\pm$ 5.76       & 0.11 $\pm$ 0.02 & 0  & --      & --       \\
NGC4605  & 5.67$_{-0.47}^{+0.47}$   & 4.53 $\pm$ 0.05   & 2.19 $_{-0.12}^{+0.12}$   & 177.77$\pm$ 6.46       & 0.09 $\pm$ 0.02 & 1  & 123.4   & 0.412    \\
NGC4941  & 1.90$_{-0.21}^{+0.21}$   & 4.06 $\pm$ 0.04   & 14.52$_{-0.71}^{+0.71}$   & 171.55$\pm$ 175.80     & 0.00 $\pm$ 0.00 & 0  & 22.1    & 0.226 &SeyfertII/(distance$>$10~Mpc)    \\
NGC5055  & 2.75$_{-0.20}^{+0.09}$   & 4.36 $\pm$ 0.09   & 14.61$_{-0.80}^{+0.80}$   & 23.07 $\pm$ 28.23      & 0.14 $\pm$ 0.06 & 1  & 107     & 0.393    \\
NGC5068  & 1.33$_{-0.23}^{+0.23}$   & 3.58 $\pm$ 0.13   & 5.18 $_{-0.46}^{+0.46}$   & 87.45 $\pm$ 15.26      & 0.23 $\pm$ 0.12 & 0  & 110.2   & 0.048    \\
NGC5194  & 0.46$_{-0.06}^{+0.06}$   & 4.41 $\pm$ 0.01   & 24.15$_{-0.53}^{+0.53}$   & 22.35 $\pm$ 2.58       & 0.26 $\pm$ 0.02 & 2  & 175.7   & 0.344 &SeyfertII/Dusty/Used IR data for modeling \\
NGC5195  & 1.27$_{-0.15}^{+0.31}$   & 4.96 $\pm$ 0.08   & 9.07 $_{-0.48}^{+2.84}$   & 117.16$\pm$ 6.54       & 0.31 $\pm$ 0.05 & 0  & 117.3   & 0.292 &Dusty/Used IR data for modeling   \\
NGC5236  & 3.46$_{-0.17}^{+1.33}$   & 5.19 $\pm$ 0.09   & 4.91 $_{-0.25}^{+0.25}$   & 135.63$\pm$ 6.22       & 0.23 $\pm$ 0.03 & 0  & 47      & 0.065  & Dusty/Used IR data for modeling\\
NGC5238  & 5.47$_{-0.51}^{+0.51}$   & 4.08 $\pm$ 0.03   & 2.15 $_{-0.08}^{+0.08}$   & 25.02 $\pm$ 3.74       & 0.09 $\pm$ 0.01 & 0  & 174.4   & 0.201    \\
NGC5457  & 2.16$_{-0.08}^{+0.08}$   & 3.78 $\pm$ 0.02   & 10.88$_{-0.31}^{+0.31}$   & 165.60$\pm$ 2.22       & 0.21 $\pm$ 0.01 & 1  & 56.4    & 0.263    \\
NGC6503  & 6.82$_{-0.66}^{+2.80}$   & 4.45 $\pm$ 0.20   & 1.62 $_{-0.39}^{+0.54}$   & 66.79 $\pm$ 84.13      & 0.00 $\pm$ 0.11 & 0  & 118.1   & 0.622  & High S\'ersic Index  \\
NGC7713  & 5.67$_{-0.49}^{+1.10}$   & 4.37 $\pm$ 0.18   & 1.80 $_{-0.23}^{+0.05}$   & 96.54 $\pm$ 3.09       & 0.37 $\pm$ 0.04 & 0  & 167.4   & 0.549    \\
\hline    
\end{tabular}
\begin{tablenotes}
$Note$: These parameters describe the NSC in the galaxy; Column 2, 3, 4, 5 and 6 give the position angle, ellipticity, S\'ersic index (n), intensity at the effective radius (I$_e$), and effective radius (R$_e$). Column 7 gives the quality of the fit to the NSC, where quality 0 is a very good fit of the model to the galaxy light with minimal errors, quality 1 is a variation of the residuals of the fit within 0.1 magnitudes. It might be caused due to the dust lanes. Quality 2 is given that had $>$0.1 magnitudes residuals that were mainly due to the obscuration of the central region with dust, and as we didn't mask the center, it affected the quality of our fits. 
\end{tablenotes}
\egroup
\end{threeparttable}

\end{table*}

\begin{table}
\centering

\caption{Multi-Gaussian Expansion (MGE) parameters of NGC~3274}
\label{table:mge}
\def\arraystretch{1.5}
\begin{threeparttable}
\begin{tabular}{ccccc}
\hline\hline

Galaxy & Filter &$I$ & $\sigma$ & q\\
&&($L_{\odot}$pc$^{-2}$) & (arcsec) &  \\ 

\hline
... & ... & ... & ... & ...\\ 
NGC~3274 & F814W &  135826    &0.007 &0.97  \\
   	 &(NSC component)& 51362.1  & 0.01  &0.97  \\
	 && 19883.1  & 0.02  &0.97 \\
     && 8916.69  & 0.04   &0.97  \\
     && 5035.47  & 0.05  &0.97  \\
     && 2337.96  & 0.09  &0.97  \\
     && 856.805  & 0.16   &0.97 \\
     && 2.65008  & 0.16   &0.97  \\
     && 203.862  & 0.34   &0.97  \\
     && 26.60  & 0.71  &0.97  \\
     &(Component 2)& 34.01  & 1.31  &0.97  \\
     && 40.41  & 2.15  &0.97  \\
     && 49.41  & 3.34  &0.97 \\
     && 54.79  & 4.99  &0.97  \\
     && 27.71  & 6.33  &0.97  \\
     && 74.83  & 8.37  &0.82  \\
     && 88.18  & 12.44  &0.82  \\
     && 84.43  & 18.23  &0.82  \\
     && 63.57  & 25.56  &0.82  \\
     && 50.38  & 33.98  &0.82  \\
     && 42.07  & 48.54  &0.82  \\
     && 12.49  & 79.38  &0.82  \\
   ... & ... & ... & ... & ...\\                
\hline

\end{tabular}
\begin{tablenotes}
$Note$: The parameters are the luminosity (I), width($\sigma$) and axis ratio(q) of each Gaussian for NGC~3274 in the F814W filter. This table is available in its entirety online.
\end{tablenotes}
\end{threeparttable}
\end{table}

\subsubsection{Profile Parameter Errors}
\par We estimated the errors in our luminosity models by using bootstrap resampling. The \texttt{IMFIT} code has the option for estimating errors using the bootstrap technique, where it generates new data by sampling pixel values with replacement from the original data image in each iteration. It then re-runs the entire fit on the new data to generate a new set of model parameters. We performed 200 iterations of bootstrapping for each galaxy, and the combined set of bootstrapped parameter values were used as a distribution for estimating confidence intervals at 68.3\% and standard deviation for each parameter. The median uncertainties in the effective radius of the NSC were found to be $\sim$10\%, while those in the S\'ersic index are $\sim$18\%. 
Table~\ref{table:sersic} includes these errors in our parameter estimates. We later propagate these errors to the MGE models for the galaxies and will propagate these error estimates into the dynamical mass estimates for the NSCs in Paper II.

\subsubsection{ Creating 1-D surface brightness profiles}

\par To visually inspect the quality of fit to the NSC in a galaxy, we created 1-D surface brightness profiles for all the galaxies. The profiles were derived using the IRAF ELLIPSE task on both the original image with mask and on model images containing the best-fit model and each component. As can be seen in Figure~\ref{fig:surf_bri} bottom panel, the NSC is seen in the inner S\'ersic component and is visible as a bump in the surface brightness profile. Here, the orange and the red lines are the 1-D profiles for individual components, whereas the black line is the combined galaxy light profile. This is compared with the data that is depicted in green crosses. The lowest panel shows the residuals in magnitudes. The PSF profile is also shown in the black dashed line.

\subsection{Constructing MGEs from the luminosity models}
\label{sec:mges}
\par To create a 3-D model of the mass distribution of the galaxy for dynamical modeling, we parametrize our best-fit models using a Multi-Gaussian Expansion (MGE) method \citep{cappellari02} that enables deprojection of the mass profile assuming axisymmetry. We need MGE models, which will be used to estimate the 3-D densities of the NSCs in Section~\ref{sec:densities} and in deriving dynamical masses using Jeans Anisotropic Models (JAM) in Paper II.  MGE models require galaxy surface brightness to be deconstructed using multiple Gaussian components with varying widths enclosing the galaxy light. Equation~\ref{eq:mge} shows the MGE projected surface brightness as:
\begin{equation}
\label{eq:mge}
\Sigma(R,\theta) = \sum_{j=1}^{N} \frac{L_j}{2\pi\sigma^2q_j} exp\left[-\frac{1}{2\sigma_j^2}\left(x^2+\frac{y_j^2}{q_j^2}\right)\right]
\end{equation}
where 
\begin{equation}
x_j = R~sin(\theta-\psi_j),~
y_j = R~cos(\theta-\psi_j)
\end{equation}
where (R,$\theta$) are the polar coordinates on the plane of the sky (x,y). N is the number of Gaussian components, L$_j$ is the total luminosity, q is the observed axial ratio, and $\sigma_j$ is the width of the Gaussian components.

From the \texttt{IMFIT} S\'ersic models, we converted individual S\'ersic components to MGEs using the \texttt{mge\_fit\_1d} code \citep{cappellari02}. This code requires the surface brightness of the galaxy to be sampled logarithmically in radius and then uses non-linear least squares minimization to fit the galaxy profile with Gaussians. We used 10 - 14 Gaussians to describe the model depending on the fit. The luminosity and the sigma ($\sigma$) of the Gaussians are then expressed in L$_\odot$pc$^{-2}$ and arcsec, respectively, which are the inputs to JAM models. As an example, Table~\ref{table:mge} gives MGEs for NGC~3274 in the F814W filter. To check the accuracy of our MGEs, we reconstructed a galaxy image from its MGEs and convolved it with its PSF. We compared the reconstructed MGE image with data and found that the uncertainty on luminosity is within 5\%.
\par To account for modeling errors, we propagated the uncertainties on \texttt{IMFIT} models to MGE models. For every iteration of the bootstrap in the \texttt{IMFIT} model, we created a corresponding MGE model and estimated the error within 1$\sigma$ for every galaxy.

\section{NSC masses and their scaling relations}
To compare with the dynamical masses we derive in Paper II; we also estimate NSC masses using their colors in this paper.  In this section, we describe our method of obtaining the colors of NSCs. Then, we use color--M/L relations similar to the method described in Section~\ref{sec:galmass} to estimate the NSC masses. Then we describe the scaling relations between NSC properties and their host galaxies. 
\label{sec:color_masses}
\subsection{Estimating colors of NSCs}

We used multi-filter $HST$ data to estimate NSC colors. The data comes from a wide variety of filters and cameras and therefore is quite heterogeneous.  Table~\ref{table:color} lists the datasets used in each galaxy to estimate the color.  For one galaxy, NGC4460, data were available in only one filter (and thus was excluded from the rest of the analysis presented here). For three other galaxies (NGC~2784, NGC~3115, NGC~3184), we do not have color information either as the image in the second filter was saturated (and they were excluded from the rest of the analysis too.)  

We estimated the magnitudes in each filter using an aperture equal to the effective radius, where the effective radius is based on the best-fitting inner Sersic profiles from Section~\ref{sec:imfit}/Table~\ref{table:sersic}. To homogenize the data, we transformed all the color information to V-I colors using a variety of transformations (see below):
\begin{enumerate}
\item WFPC2: We used the transformations as described in \citet{dolphin09}. They used the WFPC2 synthetic system was constructed using the response curves of WFPC2 instrument. Zero points and flux conversions were derived using the synthetic system for all the filters that were used to derive the photometric transformations between $HST$ filters and UBVRI, taking into account the charge transfer loss. We used the transformations from Table~4 and Equation~16 in \citet{dolphin09} and converted WFPC2's F814W, F606W, F555W, and F547M to $V-I$ colors. 

\item ACS/WFC: The conversion for ACS/WFC magnitudes is described in \citet{sirianni05}, where they used observational and synthetic transformations to determine the coefficients for transformations. The V band colors were in good agreement when determined with two methods. We used Equation~12 and Table~22 from \citet{sirianni05} for converting ACS/WFC colors to UBVRI system. We converted F606W--F814W and F555W--F814W to $V-I$ colors. 

\item WFC3/UVIS: We used the PADOVA single stellar population models\footnote{\tt \label{padova} Available from http://stev.oapd.inaf.it/cgi-bin/cmd} to convert our WFC3/UVIS magnitudes to UBVRI magnitudes. We did not use transformations described in \citet{dressel19} as it required synthetic black-body spectrum temperature for the NSC. The relationship between F555W--F814W was used to convert to $V-I$ color. 
\end{enumerate}
\begin{figure*}[ht]

\label{fig:nscmass}
\includegraphics[width=\linewidth]{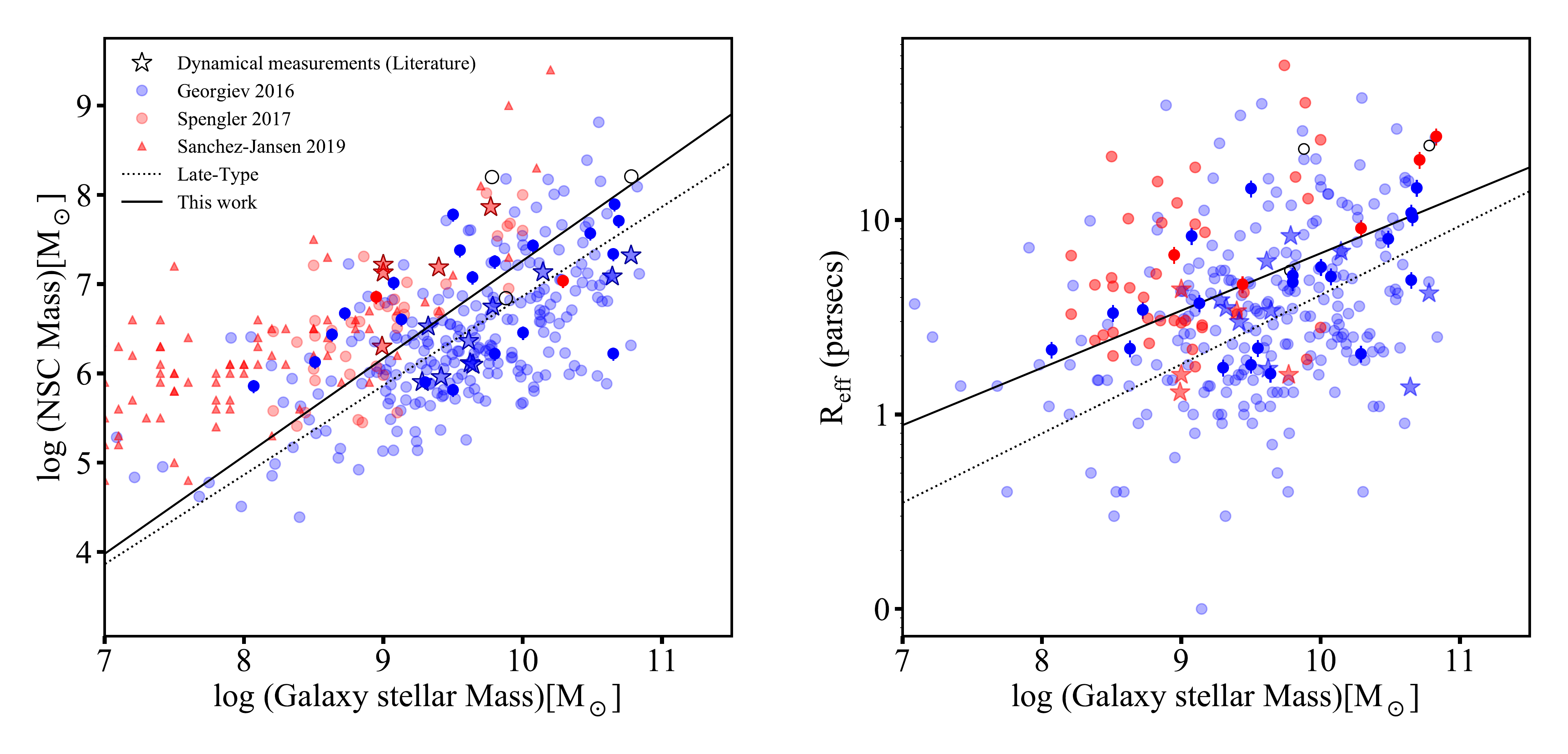}

\caption{NSC scaling relations with galaxy stellar mass.  \textit{Left:} NSC mass vs. galaxy stellar mass. \textit{Right:} NSC size vs. galaxy stellar mass.  Blue/red points indicate late-/early-type galaxies.  In both figures, the solid blue and red points are our sample galaxies, while the open circles are sample galaxies with a quality of 2 in their fits, and not included in the analyses. The NSC masses are derived using their $V-I$ colors and color--M/L relations from \citet{roediger15}.  Transparent symbols are from \citep{georgiev16,spengler17,cote06,sanchez19}.
 Gray stars are dynamical NSC mass estimates from various studies\citep{walcher05,seth10,denbrok15,nguyen18}. 
  The solid lines are the fits to the relations using our sample galaxies. The dashed lines are relationships for late-type galaxies from \citet{georgiev16}. }
\end{figure*}

\begin{figure}[ht]

\label{fig:re}
\includegraphics[width=0.5\textwidth]{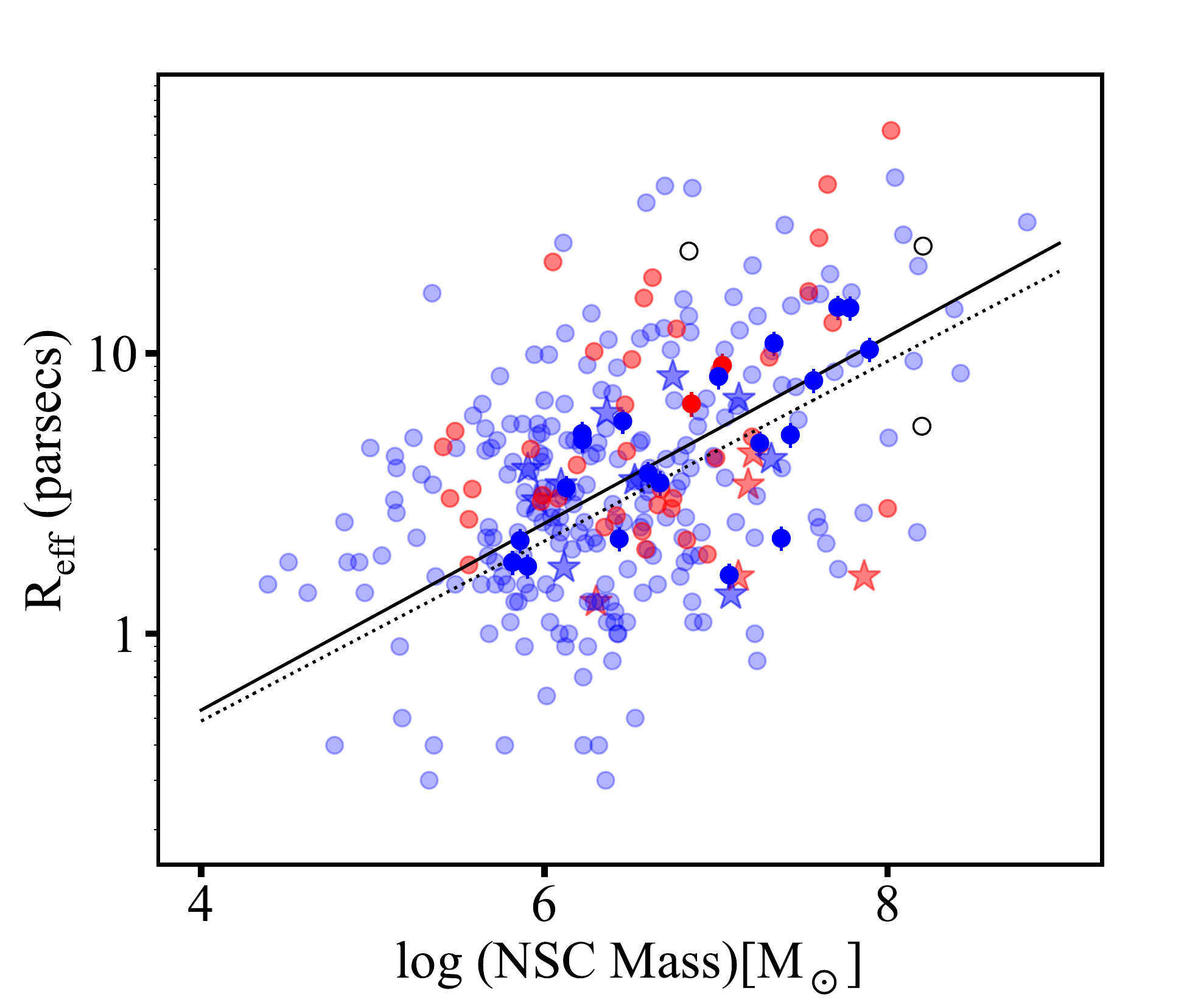}

\caption {The mass--size relation for NSCs.  Markings and data sources are the same as those in Fig.~\ref{fig:nscmass}. }

\end{figure}
\par The quoted uncertainties on the transformations for the WFPC2 and ACS transformations were 1-2\% for the V and I band filters used here. Table~\ref{table:color} shows our derived NSC colors.
\par To estimate errors on the WFC3 transformations and check the robustness of our ACS/WFPC2 color estimates, we also estimated for data from these two cameras using PADOVA single stellar population (SSP) models \citep{bressan12}\textsuperscript{\ref{padova}}. 
In some cases, the galaxies had colors redder than the unextincted colors in the models -- for these cases, we extended the models by adding additional models with extinction. Then, we used a best-fit line for transforming the colors from F555W-F814W, F547M-F814W, and F606W-F814W to $V-I$ colors. By comparing SSP model color estimates with the above colors from transformations, we found that they had a Gaussian uncertainty of $\sim$0.03 magnitudes. This color uncertainty ends up not being a significant source of error in estimating NSC masses. 

\subsection{Estimating NSC masses}
\label{sec:nscmasses}
\par We derived our NSC masses using a method similar to the one we used to derive galaxy masses in Section~\ref{sec:galmass}. Based on the colors, we estimated the masses of NSCs using the color--M/L relation from \citet{roediger15}. Since we converted all our $HST$ colors to $V-I$ colors, we used the $V-I$ vs. $I$ relation to convert those colors to $M/L$ in the $I$ band. We then apply these to the F814W-band (nearly identical to the $I$ band) luminosities to estimate the masses of the NSCs; in five cases where F814W profiles weren't used (due to lack of availability or better IR profiles), we translate their luminosity to $I$ band before determining the mass. The total luminosities of the clusters were estimated from the MGEs of the NSCs using the equation,
\begin{equation}
\label{eq:lum}
L_{tot} = \sum_{j=1}^{N} 2\pi L_j\sigma^2 q_j
\end{equation}
The galaxies that were highly extincted, we limited the M/Ls to the maximum limit from the reddest colors in existing SSP models. This is necessary because extincted red colors convert to a high $M/L$ from the color-M/L relations. We note that the intrinsic $M/L$ can still be lower than our maximum $M/L$ that we assume.

Our major source of error arises from the estimation of NSC color since the $M/L$ can vary by $\sim$25\% with a color change of 0.1 mag. To estimate the uncertainties in the NSC masses, we varied the aperture radii within which we measure the color. We changed the aperture to twice the effective radius for every galaxy and estimated the new color and corresponding NSC mass. We use the median error on all objects determined this way of $\sim$7\%, and combine this with the errors on the color--M/L relationship which we assume are 10\% \citep{roediger15}.  The errors in total luminosities from the distances and profile fits are $\sim$12.5\%. We added all these errors in quadrature to obtain a median error of 17\% on the NSC masses.

\begin{figure*}[ht]
\centering
\label{fig:ell}
\includegraphics[width=0.9\linewidth]{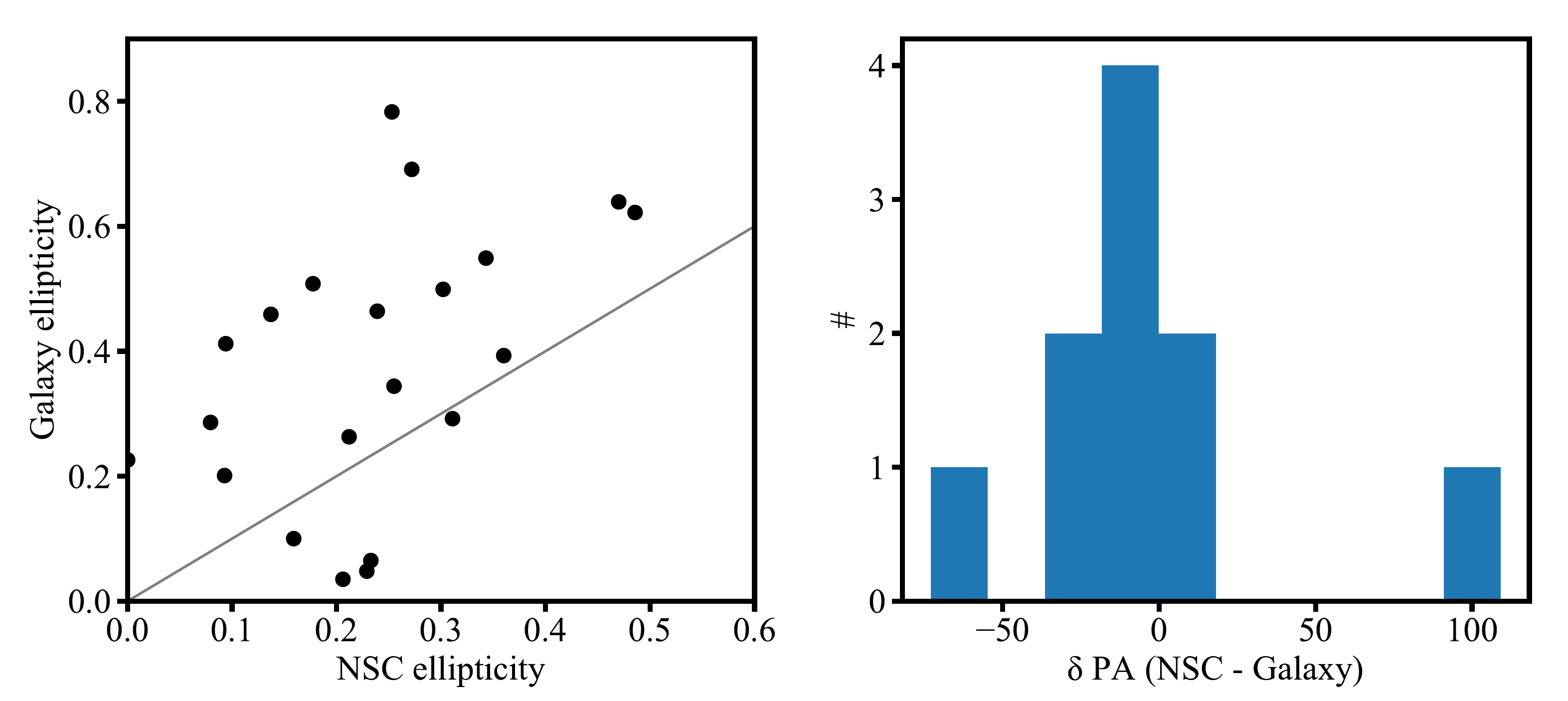}
\caption  {{\em Left}: Galaxy ellipticity vs. NSC ellipticity. A majority of  NSCs appear to be rounder than their host galaxies. The gray line shows a one-to-one correlation.  {\em Right}: The difference in position angles ($\delta$PA) between the galaxy and the NSC. Most of the NSCs have position angles within 20$^\circ$ of the galaxies' orientation. To ensure robust PA measurements, we plot only those galaxies that have both galaxy and NSC ellipticities greater than 0.2.}
\end{figure*}

\subsection{NSC masses and sizes}
\par A common approach to analyze the properties of these nuclei is to look for the possibilities of correlations that might exist between their structural parameters and the masses of the galaxies they live in. The scaling relations between NSCs and their host galaxies are an important tool for understanding the co-evolution of NSCs, SMBHs, and their hosts. Previous work has shown that NSC masses and radii are known to correlate with their host galaxy luminosities and masses \citep[e.g,][]{denbrok14,georgiev16,sanchez19}.

We plot NSC masses (given in Table~\ref{table:color}) against their host galaxy stellar mass in the left panel of Figure~\ref{fig:nscmass}. As expected, we see a correlation between the NSC mass and the galaxy mass.Since our median distance is much closer than the galaxies in \citet{georgiev16}, we can potentially provide a more accurate estimate of the relationship between galaxy mass and NSC, especially with the addition of dynamical masses in Paper II.  Our best-fit NSC mass-galaxy stellar mass relation using only our sample of galaxies (given below) is shown as the solid black line.

More than 50\% of the galaxies in our sample have NSC masses that range between 10$^6$~M$_\odot$ -- 10$^7$~M$_\odot$, reflecting the  peak in the stellar masses of our sample galaxies between $\sim$10$^9$~M$_\odot$ - 10$^{10}$~M$_\odot$. No existing dynamical mass measurements exist in galaxies below 10$^9$M$_\odot$.
While a majority of our sample are late-type galaxies, considerable data is available on the scaling of NSC masses with galaxy masses for early-type galaxies \citep[e.g.][]{cote06,turner12}.  The recent paper by \citet{sanchez19} found characteristic NSC mass of 2$\times$10$^7$~$M_\odot$ in early-type galaxies with stellar masses $\sim$5$\times$10$^9$M$_\odot$.  We include data from this study as well as multi-band stellar mass estimates from \citet{spengler17} to our plot. In the mass range probed by our data, these early-type mass estimates are similar, but clearly, a shallower slope is seen for early-type galaxies at lower masses \citep[see also][]{denbrok14}.

The right panel of Figure~\ref{fig:nscmass} shows the relation between the NSC size and its host galaxy mass.  We used the effective radius of NSC as an indicator of its size.  We can see that the NSCs grow larger as the galaxy mass increases. 75\% of our galaxies have effective radii between 1--10 pc with a median radius of the full sample of $\sim$5.18 pc. This is comparable to other studies \citep[e.g,][]{boker02,cote06}. The errors on the effective radius are determined from the bootstrapping errors on the 2-D model fits discussed in Section~\ref{sec:mges}. There are a few outliers from the relation, but as can be seen from the background gray points, similar objects were found in previous surveys on larger and more distant samples \citep{georgiev16}.  The solid relation is our best estimate and the dashed line is the late-type effective radius vs. galaxy mass relationship from \citet{georgiev16}.

Unsurprisingly given the above correlations, the size of the NSCs also correlates with NSC mass, where the more massive ones have a larger effective radius. Figure~\ref{fig:re} provides the relation between these two quantities. We also plot the relation (dashed lines) from \citet{georgiev16} for late-type galaxies. The slope of the relationship is almost identical with respect to the M$_\star$--R$_{eff}$ relation.

 The best-fit relation for correlations in  Figures~\ref{fig:nscmass} \& \ref{fig:re} are:
\begin{equation}
\label{eq:mstar_mnsc}
\rm{log}(M_{NSC}) = 1.094 * \rm log(\frac{M_{\star}}{10^6 M_\odot}) + 2.881
\end{equation}
\begin{equation}
\label{eq:mstar_reff}
\rm{log}(R_{eff}) = 0.294 * \rm log(M_{\star}) - 2.11
\end{equation}
\begin{equation}
\label{eq:mnsc_reff}
\rm{log}(R_{eff}) = 0.333 * \rm log(M_{NSC}) - 1.605
\end{equation}
We also estimated the scatter on the relationships using the method described in \citet{kelly07}. The code performs a linear regression on the data assuming an intrinsic random scatter about the data. The scatter is around $\sim$~0.13 dex for Equation~\ref{eq:mstar_mnsc}, 0.09 dex for Equation~\ref{eq:mstar_reff} and 0.06 dex for \ref{eq:mnsc_reff}.
We note that these relations exclude the galaxies in open circles whose quality of fit was not good (Qual = 2).
 
We examine the resolved 3-D density profiles of NSCs and their variation with galaxy mass further in Section~\ref{sec:density_profiles}.

\subsection{Flattening and orientation of NSCs}
In the left panel of Figure~\ref{fig:ell}, we show the ellipticity of the NSCs with respect to that of their host galaxies. We use existing host galaxy ellipticities and position angles from S4G survey \citep{sheth10}, which are available for 80\% of our sample. We observe that in general, the NSCs are rounder than their host galaxies; this is likely due to them being intrinsically rounder than the primarily late-type disks where they are hosted.  

For NSCs and galaxies with ellipticities $>$0.2, we plot the difference in position angles between the NSCs and their host galaxies in the right panel of Fig.~\ref{fig:ell}.  Although the data is limited, it provides clear evidence that NSCs a majority of these NSCs are aligned within 20$^\circ$ of their host galaxies.   A similar result was found for an edge-on galaxy sample by \citet{seth06}, who suggested that NSC and host galaxy containing similar disc alignments can result from gas accretion formation of NSCs.  On the other hand, our results seem somewhat at odds with the larger sample of late-type galaxies studied by \citet{georgievboker14}, where no similar correlation of position angles was found.  Given the relative proximity of our objects, we suggest that alignment appears common at least for galaxies with relatively flattened NSCs.

 \begin{figure}[ht]

\label{fig:sersic}
\includegraphics[width=0.5\textwidth]{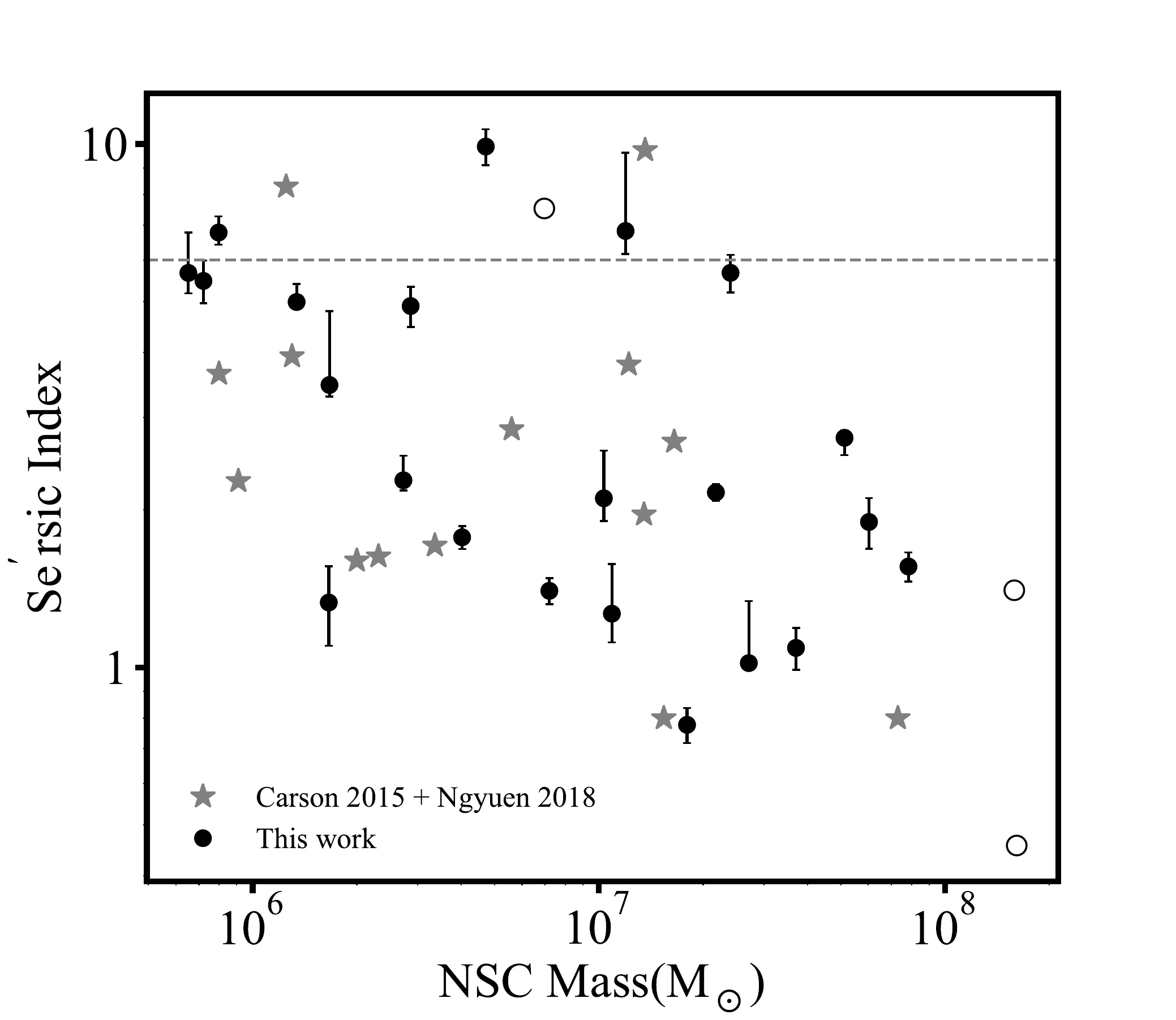}

\caption {NSC S\'ersic index vs. NSC mass. The black circles show our sample of galaxies, with open circles for the galaxies with problematic fits. The gray stars are the data from \citet{carson15} and \citet{nguyen18}. Galaxies above the gray dashed line have S\'ersic indices $>$ 6. Consistent with these previous works, we find a wide range of S\'ersic indices in our sample.}

\end{figure}

\begin{figure*}[ht]
\centering
\label{fig:density}
\includegraphics[width=0.52\linewidth]{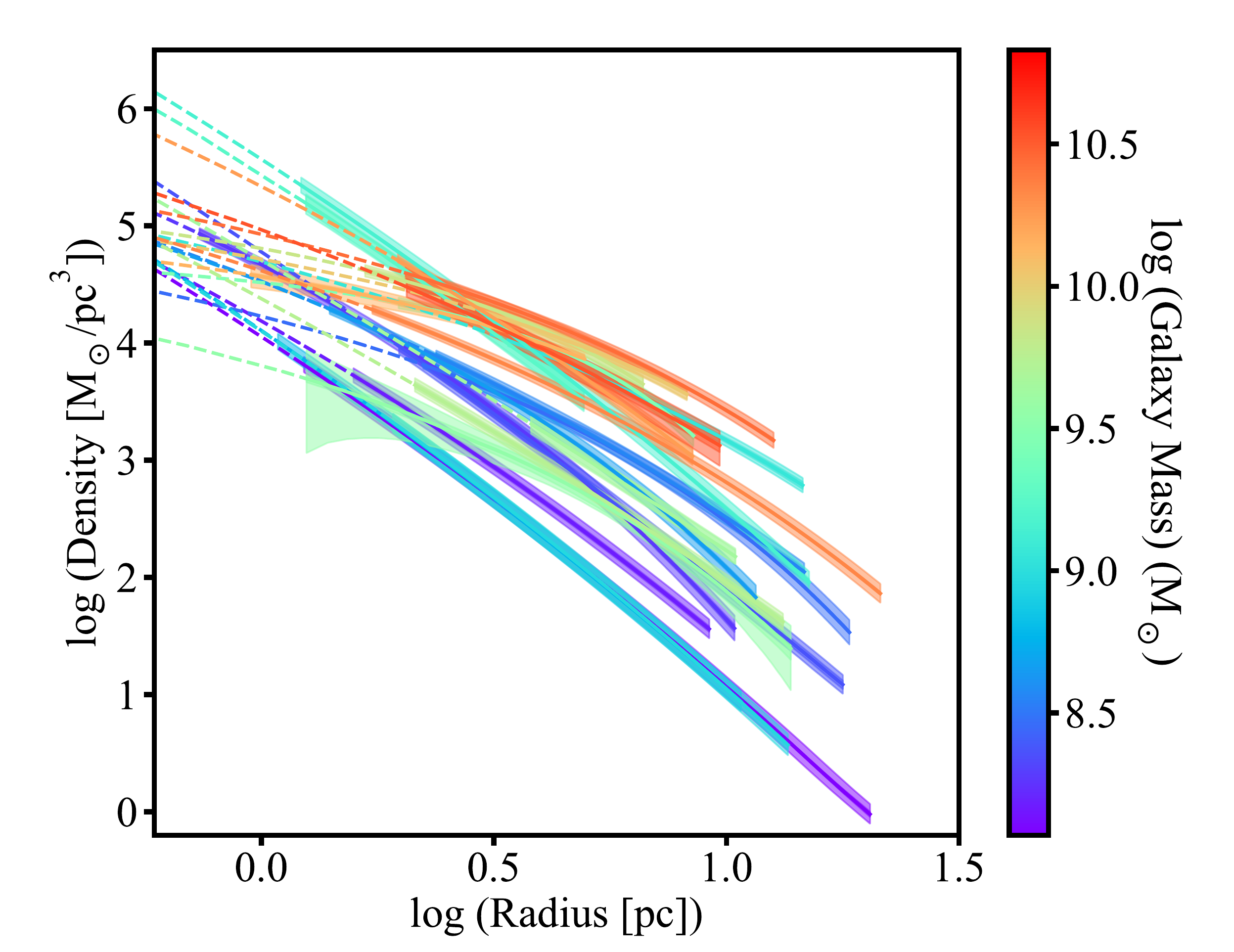}
\includegraphics[width=0.47\linewidth]{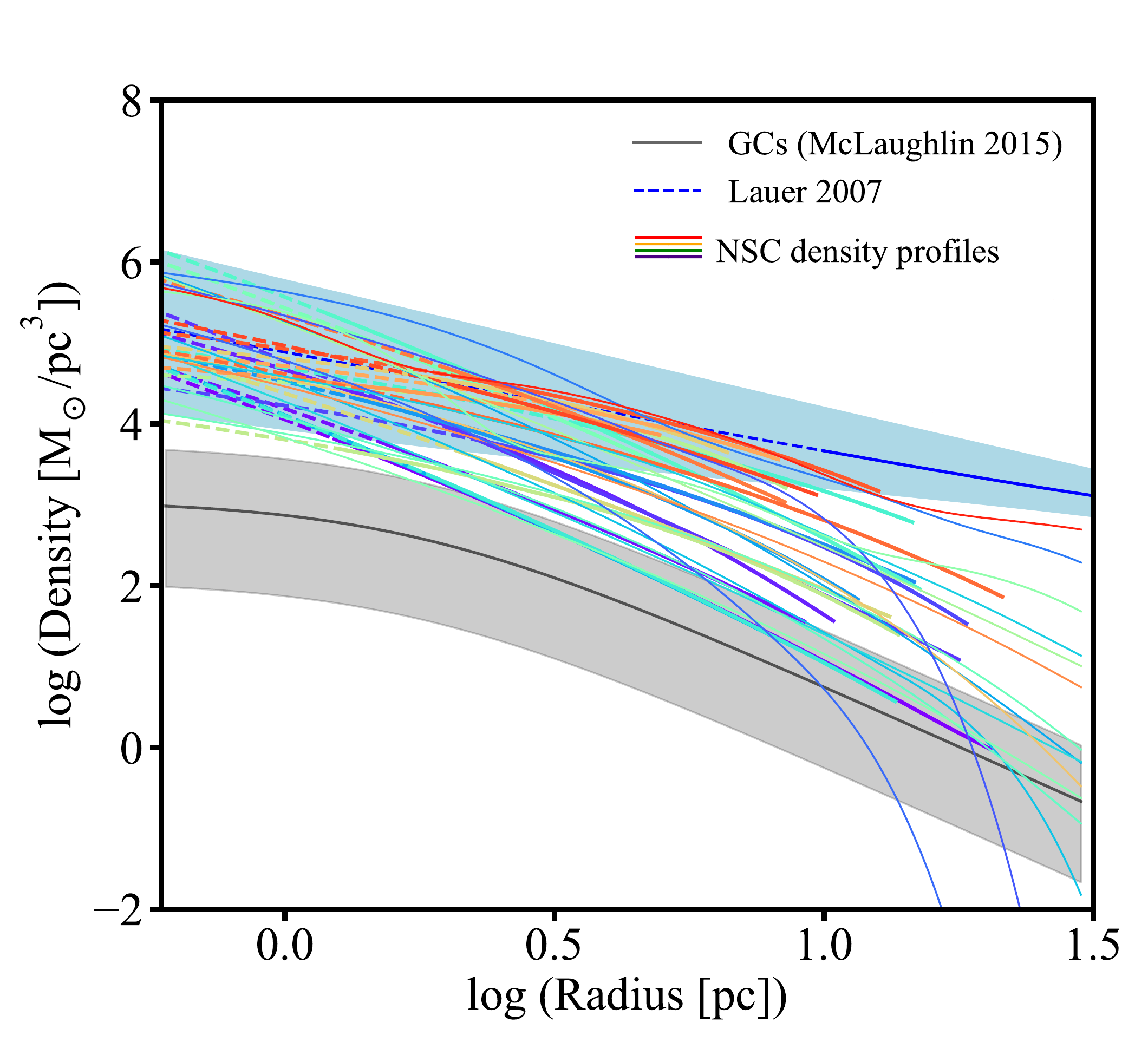}

\caption {$Left$: Deprojected 3-D density profiles of NSCs colored by galaxy stellar mass.  A correlation between galaxy mass and density is clearly seen.  These density profiles are derived in the midplane of the NSC, assuming a median inclination of 60$^\circ$.
  The NSC density profiles are plotted as solid lines where the data is well constrained.  Dashed lines indicate the inner radii where model uncertainties increase due to a lack of spatial resolution.  The data is plotted only out to the radii where the NSC falls below the surface brightness of the host galaxy.  Transparent bands show the 1$\sigma$ uncertainties in the density estimates. 
  $Right$: Our NSC density profiles combined with those from \citet{carson15} and \citet{ngyuen17}, along with Milky Way, which are also colored according to their galaxy mass. The black line is the median globular cluster density profiles from \citet{mclaughlin05} and the gray shaded region shows the interquartile range of GC profiles.  
  Similarly, the blue line is the median density profiles of galaxies from \citet{lauer07} as deprojected by \citet{stone16}, with the blue shaded region showing the interquartile range of that sample.  The blue line is dashed at radii $<$9~pc because this corresponds to a radius of 0$\farcs$05 at a median distance of the sample of 34~Mpc where the information  on the density profile becomes more uncertain.}
\end{figure*}

\subsection{S\'ersic index vs. NSC mass}
\label{sec:sersic}
\par S\'ersic indices are an indicator of the concentration of the stars in an NSC with higher values having both higher central concentrations. Figure~\ref{fig:sersic} shows the plot of S\'ersic indices of the NSCs with respect to their masses.   We do not find any clear correlation, indicating that a wide range of concentrations is possible when forming NSCs. Also, as shown in \citet{carson15}, S\'ersic indices are also wavelength dependent and can vary by about $\sim$25-30\% from a F547M filter to F814W filter. 

The wide range of S\'ersic indices observed here is consistent with the study of 10 nearby NSCs by \citet{carson15}. They concluded that the high S\'ersic indices ($>$ 6) were not due to high central concentrations but due to the behavior of profile in the wings. We have four galaxies in our sample that have S\'ersic indices $>6$; IC~5332, NGC~4242, NGC~4592, and NGC~6503. To test the cause of these high S\'ersic indices, we masked the central pixel in these galaxies and ran \texttt{IMFIT} models. We observe that the models did not change significantly even when the central pixel was masked, showing that the profile wings were dominated than the core in our fits. One possible reason for the high Sersic indices of these NSCs would be the presence of multiple components within the NSC, which when fitted with a single S\'ersic, requires high indices to explain the extended structure of the NSC.  Multiple components with distinct stellar populations and morphology are seen in several nearby NSCs \citep{seth06,seth10,nguyen18}.  The remaining galaxies had S\'ersic indices ranging from 0.5 -- 6 with a median S\'ersic index of 2.28.

\section{NSC density profiles and Tidal Disruption Events}
\label{sec:densities}
Because NSCs are the densest stellar systems in the universe, their innermost density profiles are of particular interest. This is especially true because of a recent uptick of interest in TDEs, and whose rates depend on the density of the stars around them \citep[e.g,][]{wang04,merritt09,brockamp11}. Here, we estimate the deprojected central densities and the density profiles based on the MGEs that we constructed earlier.

\subsection{Estimating NSC deprojected density profiles}
\label{sec:density_profiles}
Since we have decomposed our galaxy surface brightness to a 2-D MGE model, we can estimate its 3-D density based on each Gaussian component. Generally, the deprojection of a 2-D surface density is not unique and includes uncertainty due to the inclination and intrinsic geometry of the NSC. We use the equations for an oblate spheroid to determine the 3-D density profiles, where the ($x,y,z$) coordinates are assumed to be centered on the NSC and aligned with its principal axes. 

\begin{equation}
\label{eq:mge_density}
\rho(x,y,z) = \sum_{j=1}^{N} \frac{M_j}{(\sqrt{2\pi}\sigma_j)^3q_j} \rm{exp}\left[-\frac{1}{2\sigma_j^2}\left(x^2+y^2+\frac{z^2}{q_j^2}\right)\right]
\end{equation}
where q$_j$ is the intrinsic axial ratio, given by
\begin{equation}
q_j^2 = \frac{q_j'^2 - cos^2i}{sin^2i}
\end{equation}
and $i$ is the inclination angle. M$_j$ and $\sigma_j$ are the mass and width of each Gaussian. We get the mass from luminosity by multiplying it by the $M/L$ that we derived for each NSC using its color. We have no information about the inclination of the NSC. Therefore, we derive the densities at a median inclination for a random distribution of inclinations, $i$~=~60$^\circ$. We determine the density profiles as a function of the major axis distance ($z = 0$). Figure~\ref{fig:density} shows these density profiles. They are colored with respect to their galaxy masses and show a correlation, where the density is increasing with the galaxy mass. The primary uncertainty on these measurements are due to the distance uncertainties \ref{sec:galmass}, as well as the errors on the $M/L$ \ref{sec:nscmasses}.  In addition, we propagate the uncertainties from the bootstrapped fits to our density profiles; these are typically 7\%, and are thus smaller than the effect of the $M/L$ and distance uncertainties; however there are a couple cases where these are uncertainties are significant.

\begin{figure}[ht]
\label{fig:lumdev}
\includegraphics[width=1.02\linewidth]{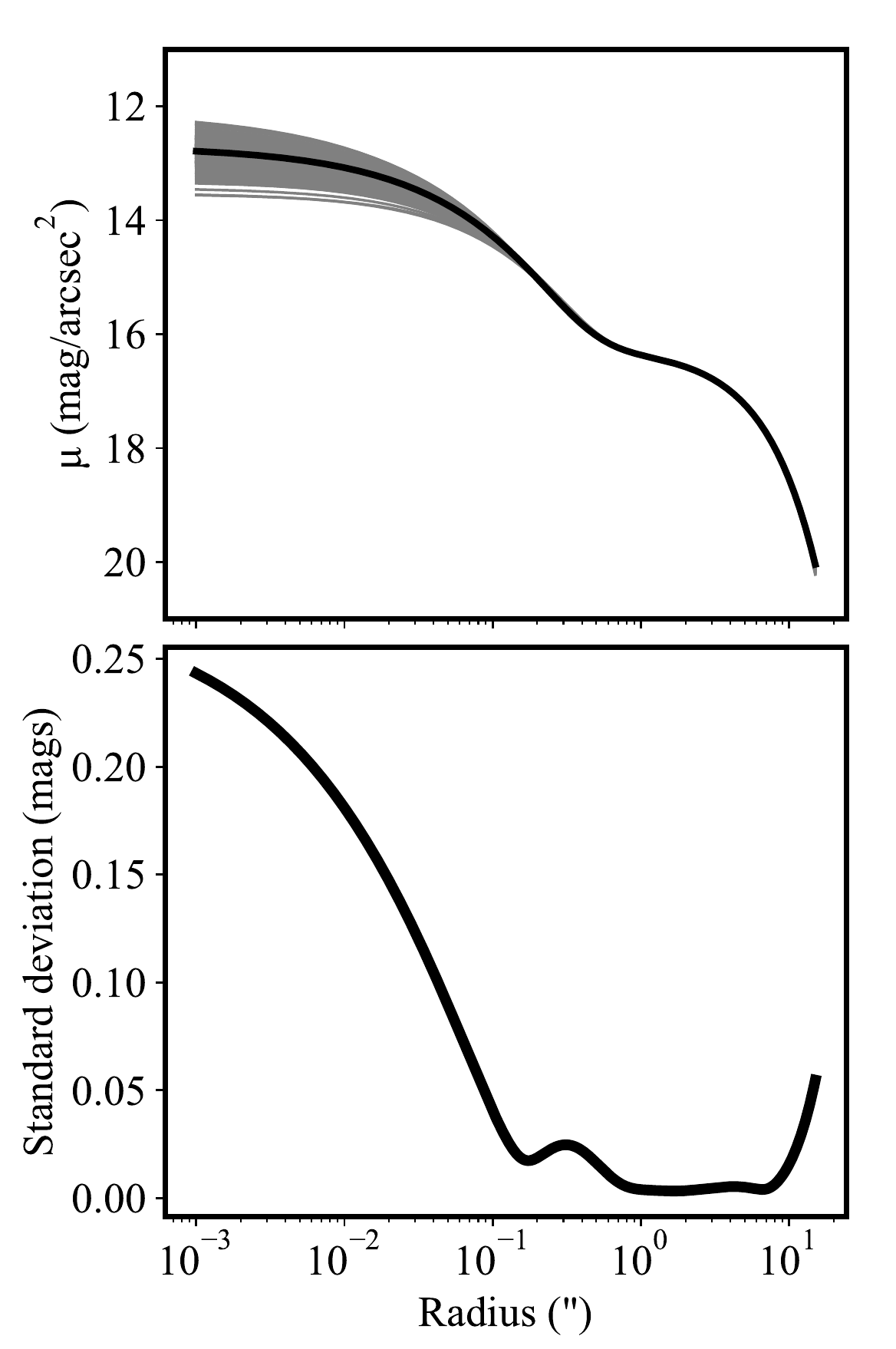}
\caption {\textit{Top}: Reconstructed surface brightness profiles of NGC~2903 from the bootstrapped {\em IMFIT} models used to estimate parameter errors. The gray lines are the different profiles from the iterations of bootstrap, with the solid black line as the best fit profile. \textit{Bottom}: The standard deviation of the varying surface brightness in the top figure. The increases towards the center are due to the lack of information at the radii $\lesssim$1 pixel in the image.  This galaxy is typical of our fits, and we use this data to constrain the radii at which our fits become less reliable.}
\end{figure}

We plot the NSC density profiles in each galaxy over the range of radii where the density profile was well constrained.  At larger radii, the NSC profiles fall below the galaxy light, and we choose this as an outer boundary.  At small radii, the HST resolution and pixel scale make the profiles uncertain. To understand this better, we estimate the surface brightness profiles for every iteration from the bootstrap model fits. The top panel of Figure~\ref{fig:lumdev} shows the variations in surface brightness profiles in NGC~2903. Since we do not have information below a certain radius, the error increases as we go towards the center of the NSC. The bottom panel of Figure~\ref{fig:lumdev} shows the plot of the standard deviation of the top panel with respect to the radius. As expected, it is close to zero where the data is available and it increases by 25\% at 0.001$''$. We make a cutoff of 10\% show the data at larger radii as a solid line in Fig.~\ref{fig:density}; this 10\% mark corresponds to a radius of $\sim$1 pixel (0$\farcs$05) in most galaxies. At smaller radii, we plot the inward extrapolation of the best fit S\'ersic profiles as dashed lines.  We also derived an outer limit to the radius based on where the galaxy luminosity equals the NSC luminosity. Beyond this radius, the profile of the NSC again becomes more model-dependent and is based on extrapolation of the assumed light profile to radii where the data can't be constrained. We plot all the galaxies that had a quality of 0 and 1.

The right panel of Figure~\ref{fig:density} shows additional data from different studies. The gray shaded region shows typical densities of globular clusters (GCs) from \citet{mclaughlin05}, where they fit 153 GCs to derive their structural and dynamical properties. The black line is the median GC 3-D density using the King profiles, and the shaded region was derived using quartiles for the GC distribution.
The GCs typically have lower densities than the NSCs and have flatter profiles near the center compared to the NSCs. We extrapolate the densities of NSCs lower than their resolution and plot them as the dashed lines. We note that \citet{mclaughlin05} GC profiles are derived from fitting modified King (Wilson) profiles to the Galactic GCs, while in this paper, we use a S\'ersic, which might result in slightly different extrapolation toward the center. The blue region is a representative of 3-D density profiles from \citet{stone16} of 167 galaxies. These galaxies are typically further in distance than our sample, and therefore most of their profiles are extrapolations below 10~pc. The dashed blue line is the median of those density profiles. 
We also include density profiles for other samples of nearby galaxies (thin colored lines) including 10 late-type galaxies from \citet{carson15} (with mass-to-light ratios estimated using color--$M/L$ relations) and dynamical masses for five early-type galaxies from \citet{ngyuen17,nguyen18}. We also include the density profile of Milky Way using the MGEs and $M/L$ from \citet{feldmeierkrause17}.

Fig.~\ref{fig:density} clearly shows a correlation between the density profiles and the galaxy stellar mass, with higher mass galaxies having NSCs with higher densities at all radii.  These higher mass NSCs also appear to have flatter density profiles.  We quantify these effects below.  The right-hand panel shows that the NSCs fall above the density of typical MW GCs at all radii, a fact we discuss more in Section~\ref{sec:implications}.
Compared to the density profiles of high mass ellipticals presented by \citet{lauer07}, the central densities are similar, but our NSC profiles appear to drop off more quickly.  We note that the a typical galaxy in the \citet{lauer07} sample is more massive ($\sim$10$^{11}$~M$_\odot$) than any of the galaxies we consider here.  This trend towards higher mass galaxies having flatter and higher density profiles is consistent with the mass trend we see in our galaxies.

\par To visualize Figure~\ref{fig:density} in another way, we plot the densities of the galaxies at a fixed physical radius in the top panel of Figure~\ref{fig:med_density}. We choose to plot the density at a radius of 5~pc corresponding to the median effective radius of the NSCs in our sample and a radius at which all the NSCs density profiles are well constrained.   We observe that the densities increase with respect to the galaxy mass at a fixed radius, although the scatter also seems to increase with increasing mass. The estimates will be more accurate after we perform the dynamical mass measurements for the NSCs in Paper II. We derive a correlation between the galaxy mass and the density of the NSC at 5~pc. We used a similar technique employed earlier to quantify the correlation and scatter using the method from \citet{kelly07}. We estimate an intrinsic scatter of 0.28 dex on this correlation and find: 
\begin{equation}
\label{eq:density}
\rm{log}(\rho_{\rm 5 pc}) = 0.592*\rm{log}(M_\star/10^9M_\odot)+2.819
\end{equation}

To aid modelers in creating realistic nuclear mass profiles, we also estimate the power-law mass dependence $\rho \propto r^\gamma$ over the region we have fitted in each object.  The best fit value for each galaxy is shown in the bottom panel of Figure~\ref{fig:med_density}.  We find a median $\gamma$ of -2.10,
with a standard deviation of 0.58.
As above, we derive a relation between the power-law index ($\gamma$) and the galaxy mass.  
\begin{equation}
\label{eq:gamma}
\gamma = 0.963*\rm{log}(M_\star/10^9M_\odot)-2.762 
\end{equation}
with a  scatter of 0.32 dex.  Combined with the galaxy mass dependence and scatter of the densities at 5 pc, this should enable accurate predictions of nuclear density profiles for galaxies in the mass range between 3$\times$10$^8$ and 8$\times$10$^{10}$ M$_\odot$ and at radii between $\sim$1-20 pc.

\subsection{Implications of Density Profiles for NSC formation}
\label{sec:implications}
These densities provide an important insight into the formation scenarios of NSCs. Two main formation mechanisms exist for NSCs: (1)~$In$~$situ$ formation, where the NSCs form because of the gas falling into the center of a galaxy \citep{mclaughlin06,hopkins10,brown18}. Repeated episodes of star formation are evidence of this type of formation \citep{walcher06,kacharov18}. (2)~Cluster accretion, where star clusters merge and sink into the center due to dynamical friction. For example, \citet{gnedin14} suggests that at low galaxy masses $M_{NSC}$~$\propto$~$M_{gal}^{0.5}$ that is an indicator of cluster inspiral, and this slope is indeed found in low-mass early-type galaxies \citep{denbrok14,sanchez19}. Simulations \citep{hartmann11,antonini15,cole17} have hinted that the formation of NSCs can be formed from both the processes.

\begin{figure}[ht]

\label{fig:med_density}
\includegraphics[width=1.02\linewidth]{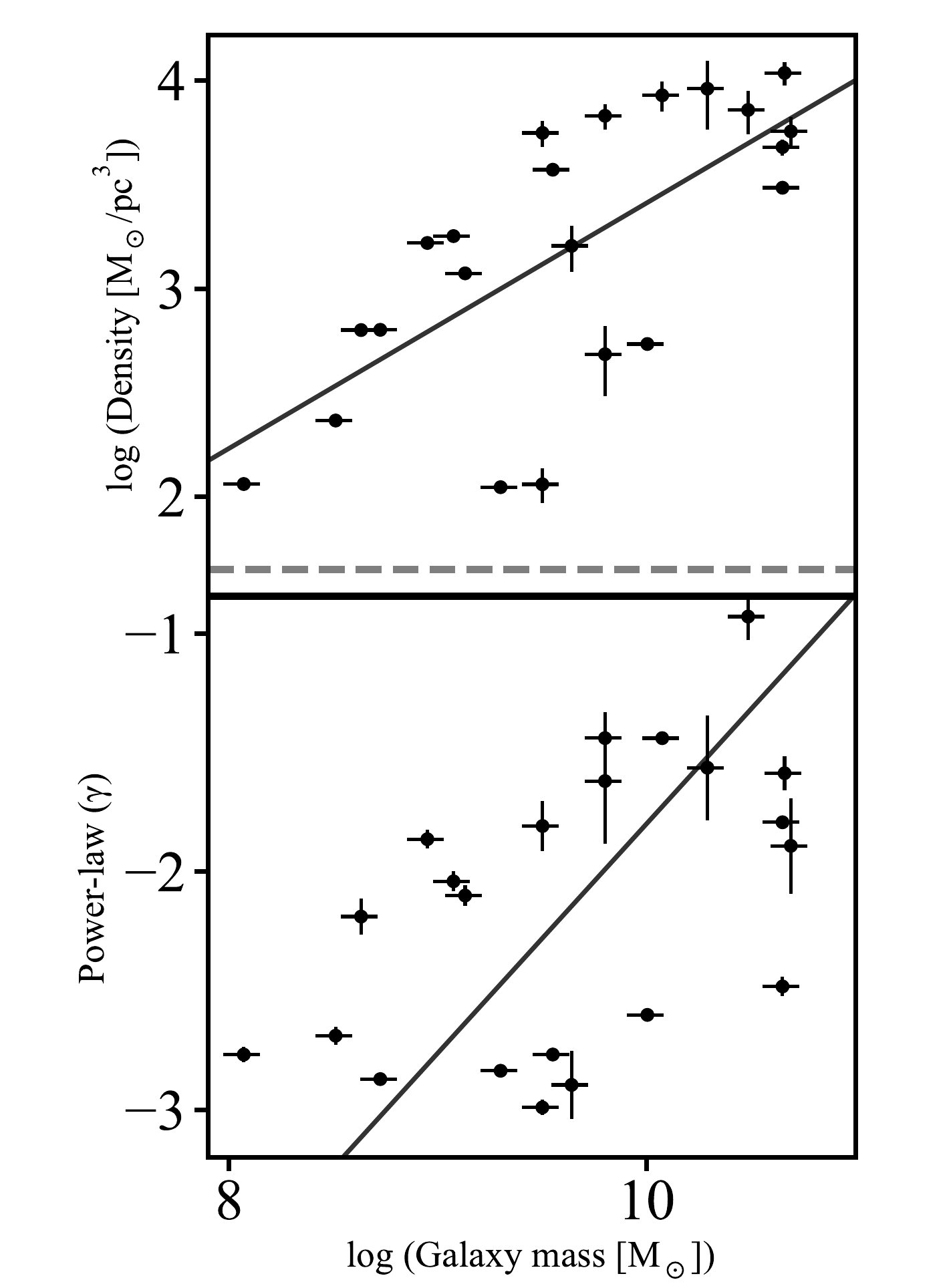}

\caption  {$Top$: Densities of NSCs at a median effective radius of 5~pc. The solid line gives the relation between galaxy mass and deprojected density. The dashed line is the typical median density of Milky Way globular clusters at a radius of 5~pc from the \citet{mclaughlin05} sample. $Bottom$: Power-law index ($\gamma$) of the NSC density profiles. The density profiles are typically steeper at low-masses and get flatter as the galaxy mass increases.}
\end{figure}

In Figure~\ref{fig:med_density} the gray dashed line is the typical density of a globular cluster at 5~pc.  All the NSCs are much denser than the typical GCs, and the density increases with galaxy mass. Our density profiles provide two clues about the formation of NSCs.  First of all, in merging systems, the slope of the density profiles reflects the steepest slope of the merged systems \citet{dehnen05}, thus the fact that the profiles of most of the NSCs appear steeper than that of typical globular clusters may suggest an important role for {\em in situ} star formation.  Second, in simulations of merging clusters, the merger product is typically similar or lower density than the original systems \citep[e.g.][]{hartmann11,antonini11}.  The high density of the higher mass NSCs may, therefore, suggest a preferential role for {\em in situ} star formation, especially at the highest masses. However, we also note that the existing Milky Way globular cluster system may not be fully representative of the clusters that merge to form the NSC, for instance, the Arches cluster near the Galactic center is one of the densest clusters in the Milky Way with a central density of $\sim$10$^5$~M$_\odot$/pc$^3$ \citep{portegies10}.

\subsection{Connection to Tidal Disruption Events}

TDEs are rare events, and, therefore the modeling of TDE rates is vital for understanding the demographics of the galaxies in which they occur.  
The central densities of NSCs thus, can be useful in the prediction of the TDE rates as they are one of the primary ingredients in rate models \citep[e.g.][]{stone16}. The majority of TDEs are also known to occur in galaxies with masses ranging from 10$^9$ - 10$^{11}$ M$_\odot$, with no TDEs observed in galaxies with masses $<$~10$^9$M$_\odot$ \citep{vanvelzen14,lawsmith17,graur18}. Our data probes this mass range, and our  density profiles can, therefore, be useful in predictions of TDE rates.

Theoretical predictions suggest the rates of these TDEs to be 10$^{-4}$--10$^{-5}$~year$^{-1}$ galaxy \citep{magorrian99,wang04,stone16}. Some observations have suggested a similar rate \citep{esquej08,auchettl18}, while others have suggested lower rates \citep{vanvelzen14,holoien16b}. The current observations of TDEs are small ($\sim$40), and the inconsistency in these estimates is likely at least in part due to this limitation.

Future surveys like Pan-STARRS, LSST, PTF will be able to detect more of these events, and we will be able to constrain these rates more accurately using observations.  Combining these observations with more sophisticated rate modeling will constrain the demographics of black holes in galaxies  \citep{stone16,vanvelzen18}.
 
The existing rate predictions in \citet{stone16} are based on density profiles of early-type galaxies derived by \citet{lauer07}; these are shown as a blue band in Fig.~\ref{fig:density}.  While the central densities appear to be comparable, our data represent an improvement in several respects with regards to TDE modeling: (1) the Lauer sample is dominated by massive galaxies, and focuses on early-type galaxies where TDEs are rarely found \citep[e.g.][]{lawsmith17}, (2) the median distance of the Lauer sample is $34$~Mpc, with only a single galaxy $<$10~Mpc, thus the physical resolution of the sample is lower, and (3) the Lauer sample excludes the presence of NSCs in fitting its density profiles, while a majority of the galaxies in the mass range of typical TDEs do host NSCs \citep[e.g.][]{georgievboker14,sanchez19}.  We, therefore, suggest that using our galaxy mass -- density and galaxy mass -- power-law index relations will improve the accuracy of existing TDE rate predictions.  We will improve these relations with the addition of dynamical mass estimates in our next paper.

\section{Summary}
\label{sec:conclusion}
In this work, we analyzed a sample of 29 nearby galaxies that are a volume-limited heterogeneous sample within 10 Mpc.

We summarize our main results here:
\begin{enumerate}

  \item We compile  galaxy stellar masses using both color-$M/L$ relations \citep{roediger15} and IR luminosity measurements from the Spitzer Local Volume Legacy survey \citet{cook14}.

  \item We used high-resolution $HST$ data to create \texttt{IMFIT} models \citep{erwin15} to parameterize the NSC morphology and the central regions of their host galaxies.   We present S\'ersic profile fits to the NSCs derived from data in the reddest filter available. These models will be used to calculate their dynamical masses and densities using the kinematic data available from GNIRS and XSHOOTER in the Paper II. 

   \item We derive NSC masses using luminosities from our model fits combined with $HST$ based color measurements.

   \item We present an NSC mass - galaxy mass correlation, and an effective radius - galaxy mass correlation. Because of the proximity of our sample, these data are based on higher resolution data than previous works \citep[e.g.][]{georgievboker14,georgiev16}. Half of our NSC masses lie between 10$^6$--10$^7$~M$_\odot$.
     
   \item We compare the flattening and orientation of NSCs to their host galaxies, and find that typically, NSCs are rounder than their host galaxies.  Most flattened NSCs position angles are aligned with their host galaxies, favoring {\em in situ} formation as the dominant mechanism for these NSCs. 
     
   \item We present the first large sample of measurements of 3-D stellar density profiles of NSCs on scales down to $\sim$1~pc.  These profiles show that the NSC densities and density slopes are correlated with galaxy mass.  We derive empirical relations for the galaxy mass dependence of both the central density and the power-law slope of the density profile.

   \item  NSCs have densities higher than typical GCs, especially in the highest mass galaxies.  This may suggest that these densest NSCs are created primarily by star formation within the NSC, and not through dynamical inspiral of GCs.   

   \item We discuss the implications of our derived NSC density profiles for estimating TDE rates.  Our galaxy sample matches the mass range of a majority of TDE hosts and has density information closer to the center than previous density profiles used for estimating TDE rates \citep[e.g.][]{stone16}.  We, therefore, suggest that the use of our galaxy mass-density relations can improve TDE rate estimates.
     
\end{enumerate}
{\em Acknowledgments:} The authors thank Nicholas Stone for sharing his density profile data with us, and Joel Roediger for sharing his models.  We also acknowledge helpful conversations with Linda Strubbe, Monica Valluri, Fabio Antonini, Alessandra Mastrobuono-Batisti, and Sjoert van Velzen.  RP and ACS acknowledge support from NSF AST-1350389.

\bibliography{enhanced_ml}

\setcounter{table}{0}
\renewcommand{\thetable}{A\arabic{table}}
\appendix 
 \label{sec:appendix}

\begin{table}[]
\centering
\caption{Galaxy Sample Properties}
\label{table:params}
\begin{threeparttable}
\bgroup
\def\arraystretch{1.6}
\setlength\tabcolsep{10pt}
\begin{tabular}{lllllllll}
 \hline\hline
Galaxy   & Distance (Mpc) & T-type & A$_v$    & m$_K$     & M$_B$    & $g-i$  & $B-V $ & M$_\star $(M$_\odot$)    \\
(1) & (2) & (3) & (4) & (5) & (6) &(7)& (8) & (9)\\
\hline
Circinus & 4.207    & 3      & 2.1   & 4.984  & -15.1 & -    & 0.75 & 3.06$\times$10$^{10}$ \footnote{\label{bmv} Stellar ass derived from $B-V$ color}\\
ESO274-1 & 2.79     & 7      & 0.691 & 8.357  & -15.8 & -    & 0.64 & 4.28$\times$10$^{8}$ \footnoteref{bmv}\\
IC5052   & 5.49     & 7      & 0.139 & 8.884  & -17.3 & -    & 0.4  & 1.35$\times$10$^{9}$ \footnote{\label{cook} Stellar mass from \citet{cook14}}\\
IC5332   & 7.79     & 7      & 0.046 & 8.704  & -18.5 & -    & 0.5  & 7.59$\times$10$^{9}$ \footnoteref{cook}\\
NGC2784  & 9.81     & -2     & 0.575 & 6.322  & -18.8 & -    & 0.86 & 5.13$\times$10$^{10}$ \footnoteref{bmv}\\
NGC2787  & 7.48     & 1      & 0.365 & 7.263  & -17.5 & -    & 0.84 & 1.19$\times$10$^{10}$ \footnoteref{bmv}\\
NGC2903  & 8.87     & 4      & 0.085 & 6.036  & -20.3 & 0.99 & 0.58 & 4.57$\times$10$^{10}$ \footnoteref{cook}\\
NGC3115  & 9.68     & -3     & 0.127 & 5.883  & -20.1 & -    & 0.83 & 6.76$\times$10$^{10}$\footnoteref{bmv}\\
NGC3115B & 9.7      & -2     & 0.144 & 10.053 & -16.3 & -    & 0.57 & 8.90$\times$10$^{8}$ \footnoteref{bmv}\\
NGC3184  & 11.11    & 6      & 0.046 & 7.225  & -19.3 & 1.07 & 0.56 & 1.95$\times$10$^{10}$ \footnote{\label{gmi} Stellar mass derived from $g-i$ color}\\
NGC3274  & 7.98     & 6      & 0.066 & 10.742 & -15.8 & 0.4  & 0.27 & 3.24$\times$10$^{8}$ \footnoteref{cook}\\
NGC3344  & 9.82     & 4      & 0.091 & 7.437  & -18.8 & 0.94 & 0.53 & 6.31$\times$10$^{9}$ \footnoteref{cook}\\
NGC3593  & 10.81    & 0      & 0.053 & 7.417  & -17.3 & 1.22 & 0.84 & 6.03$\times$10$^{9}$ \footnoteref{cook}\\
NGC4242  & 5.27     & 8      & 0.033 & 9.116  & -17.9 & 0.65 & 0.48 & 2.00$\times$10$^{9}$ \footnoteref{cook}\\
NGC4460  & 9.2      & -1     & 0.052 & 9.056  & -17.6 & 0.85 & -    & 2.75$\times$10$^{9}$ \footnoteref{cook}\\
NGC4517  & 8.36     & 6      & 0.065 & 7.329  & -19.6 & 1.07 & 0.63 & 1.01$\times$10$^{10}$ \footnoteref{gmi}\\
NGC4592  & 9.6      & 8      & 0.061 & 10.218 & -18.5 & 0.59 & -    & 5.30$\times$10$^{8}$ \footnoteref{gmi}\\
NGC4600  & 9.29     & 1      & 0.073 & 9.814  & -15.6 & 1.02 & -    & 1.19$\times$10$^{9}$ \footnoteref{gmi}\\
NGC4605  & 5.55     & 8      & 0.039 & 7.762  & -17.8 & 0.69 & 0.37 & 3.55$\times$10$^{9}$ \footnoteref{cook}\\
NGC4941  & 15       & 2      & 0.099 & 8.217  & -17.3 & -    & 0.79 & 3.17$\times$10$^{9}$ \footnoteref{bmv}\\
NGC5055  & 9.04     & 4      & 0.048 & 5.608  & -20.1 & 1.1  & 0.67 & 4.90$\times$10$^{10}$ \footnoteref{cook}\\
NGC5068  & 5.45     & 6      & 0.281 & 7.549  & -18.6 & -    & 0.44 & 6.31$\times$10$^{9}$ \footnoteref{cook}\\
NGC5194  & 8.39     & 5      & 0.096 & 5.496  & -20.9 & 0.95 & 0.63 & 6.03$\times$10$^{10}$ \footnoteref{cook}\\
NGC5195  & 7.66     & -1     & 0.097 & 6.251  & -19.1 & -    & 0.85 & 1.95$\times$10$^{10}$ \footnoteref{cook}\\
NGC5236  & 4.89     & 5      & 0.182 & 4.619  & -20.1 & -    & 0.55 & 4.47$\times$10$^{10}$\footnoteref{cook} \\
NGC5238  & 4.51     & 8      & 0.027 & 14.725 & -15   & 0.73 & 0    & 1.17$\times$10$^{8}$ \footnoteref{cook}\\
NGC5457  & 6.95     & 6      & 0.023 & 5.512  & -21.1 & 0.93 & 0.45 & 4.47$\times$10$^{10}$ \footnoteref{cook}\\
NGC6503  & 6.28     & 6      & 0.088 & 7.296  & -17.6 & -    & 0.61 & 4.37$\times$10$^{9}$ \footnoteref{cook}\\
NGC7713  & 7.79     & 6      & 0.045 & 9.195  & -18.5 & -    & 0.28 & 3.16$\times$10$^{9}$\footnoteref{cook}\\
\hline
\end{tabular}
\begin{tablenotes}
{\it List of parameters for the sample of 29 galaxies} (1) Galaxy name as described by Hyperleda (2) Distance to the galaxy adopted from \citet{karachentsev04} updated catalog (3) Hubble type of the galaxy (4) Extinction of the galaxy in V band from \textit{Nasa Extragalactic Database} (5) K-band absolute magnitude from \citet{karachentsev04} (6) B-band absolute magnitude from \citet{karachentsev04} (7) $g-i $colors from the NASA Sloan ATLAS (8) Effective $B-V$ colors from Hyperleda database (9) Stellar masses of the galaxy as derived using colors described in Section~\ref{sec:galmass}. The errors on the masses are of the order of $\sim$30\%.
\end{tablenotes}
\egroup
\end{threeparttable}
\end{table}

\newpage
\begin{table}
\centering
\caption{{\em HST} observations used}
\label{table:HST}
\begin{threeparttable}
\bgroup
\def\arraystretch{1.6}
\setlength\tabcolsep{10pt}
\begin{tabular}{ccccc}
\hline\hline
Galaxy & PropId & Camera    & Filter & Exptime \\
(1) & (2) & (3) & (4) & (5) \\
\hline 
Circinus & 7273   & WFPC2/PC  & F814W  & 40     \\
ESO274-1 & 10235  & ACS/WFC   & F814W  & 900    \\
IC5052   & 9765   & ACS/WFC   & F814W  & 700    \\
IC5332   & 9042   & WFPC2/WF3  & F814W  & 460    \\
NGC2784  & 8591   & WFPC2/PC  & F547M  & 1600   \\
NGC2787  & 6633   & WFPC2/PC  & F814W  & 730    \\
NGC2903  & 5211   & WFPC2/PC  & F814W  & 820    \\
NGC3115  & 5512   & WFPC2/PC  & F555W  & 1330   \\
NGC3115B & 5999   & WFPC2/PC  & F814W  & 320    \\
NGC3184  & 13350   & WFC3/UVIS & F606W  & 181    \\
NGC3274  & 13364  & WFC3/UVIS & F814W  & 980    \\
NGC3344  & 13364  & WFC3/UVIS & F814W  & 980    \\
NGC3593  & 11128  & WFPC2/PC  & F814W  & 260    \\
NGC4242  & 13364  & WFC3/UVIS & F814W  & 986    \\
NGC4460  & 5446   & WFPC2/PC  & F606W  & 160    \\
NGC4517  & 9765   & ACS/WFC   & F814W  & 700    \\
NGC4592  & 11360  & WFC3/UVIS & F555W  & 1044   \\
NGC4600  & 12878  & ACS/WFC   & F814W  & 2076   \\
NGC4605  & 13364  & WFC3/UVIS & F814W  & 1001   \\
NGC4941  & 8597   & WFPC2/PC  & F606W  & 560    \\
NGC5055  & 8591   & WFPC2/PC  & F547M  & 1400   \\
NGC5068  & 8599   & WFPC2/PC  & F814W  & 640    \\
NGC5194  & 12490  & WFC3/IR   & F110W  & 611    \\
NGC5195  & 12490  & WFC3/IR   & F110W  & 611    \\
NGC5236  & 11360  & WFC3/IR   & F160W  & 2396   \\
NGC5238  & 10905  & ACS/WFC   & F814W  & 1158   \\
NGC5457  & 9490   & ACS/WFC   & F814W  & 720    \\
NGC6503  & 13364  & WFC3/UVIS & F814W  & 1016   \\
NGC7713  & 9892   & ACS/WFC  & F625W  & 200    \\
\hline
\end{tabular}
\begin{tablenotes}
{\it Data used for modeling (from $HST$ archive)}: (1)  Galaxy name as described by Hyperleda  (2) Proposal ID from which the data was used. Column (3) and (4) are the camera and filter that was used to derive the luminosity models. Column(5) is the total exposure time that was available for the data.
\end{tablenotes}
\egroup
\end{threeparttable}
\end{table}
\newpage

\begin{table}[]
\centering
\caption{Additional NSC properties}
\label{table:color}
\begin{threeparttable}
\bgroup
\def\arraystretch{1.8}
\setlength\tabcolsep{6pt}
\begin{tabular}{ccccccccccc}
\hline \hline
Galaxy   & Camera     & Filter1 & Filter2 & Color  & Color (V-I) & M/L$_I$  & Luminosity & NSC mass & log($\rho_{5pc}$) &  $\gamma$ \\
& & & & (Filt2 - Filt1)& & (M$_\odot$/L$_\odot$)& (L$_\odot$)&(M$_\odot$)& (M$_\odot$pc$^{-3}$) & \\
(1) & (2) & (3) & (4) & (5) & (6)& (7) & (8) & (9) & (10) & (11)\\
\hline
Circinus & WFPC2      & F814W   & F547M   & 0.88   & 1.2    & 5.48   & 6.77$\times$10$^6$     & 3.71$\times$$10^7$ & 2.31      & -2.77   \\
ESO274-1 & ACS/WFC    & F814W   & F606W   & 1.07   & 1.44   & 2.11   & 1.29$\times$10$^6$     & 2.73$\times$$10^6$ & 2.21      & -2.69   \\
IC5052   & ACS/WFC    & F814W   & F606W   & 0.52   & 0.71   & 4.53   & 8.89$\times$10$^5$     & 4.03$\times$$10^6$ & 2.43      & -2.19   \\
IC5332   & WFPC2      & F814W   & F606W   & 0.81   & 1.1    & 5.48   & 1.27$\times$10$^6$     & 6.96$\times$$10^6$ & 2.60      & -2.87   \\
NGC2784  & WFPC2      & F606W   & -       & -      & -      &	-    & 1.78$\times$10$^7$	    & - 		& -         & -    	\\
NGC2787  & WFPC2      & F814W   & F555W   & 0.5    & 0.68   & 3.43   & 7.90$\times$10$^6$     & 2.71$\times$$10^7$ & 3.04      & -1.86   \\
NGC2903  & WFPC2      & F814W   & F555W   & 1.18   & 1.15   & 5.48   & 1.43$\times$10$^7$     & 7.83$\times$$10^7$ & 3.00      & -2.04   \\
NGC3115  & WFPC2      & F555W   & -       & -      & -      & -      & 1.44$\times$10$^8$	  & - 			& -         & -    	\\
NGC3115B & WFPC2      & F814W   & F555W   & 1.11   & 1.07   & 1.36   & 5.28$\times$10$^6$     & 7.19$\times$$10^6$ & 2.37      & -2.10   \\
NGC3184  & WFC3/UVIS  & F606W   & -       & -      & -      & -      & 6.91$\times$10$^5$	  & -			& -         & -    	\\
NGC3274  & WFC3/UVIS  & F814W   & F555W   & 1.57   & 1.53   & 1.26   & 1.06$\times$10$^6$     & 1.34$\times$$10^6$ & 1.97      & -2.84   \\
NGC3344  & WFC3/UVIS  & F814W   & F555W   & 1.04   & 1.02   & 0.68   & 2.64$\times$10$^7$     & 1.80$\times$$10^7$ & 2.50      & -2.99   \\
NGC3593  & WFPC2      & F814W   & F547M   & 0.83   & 0.8    & 5.48   & 2.89$\times$10$^7$     & 1.58$\times$$10^8$ & -         & -       \\
NGC4242  & WFC3/UVIS  & F814W   & F555W   & 0.99   & 0.95   & 1.05   & 7.63$\times$10$^5$     & 7.98$\times$$10^5$ & 3.51      & -1.81   \\
NGC4460  & WFPC2      & F606W   & -       & -      & -      &	-    & 7.17$\times$10$^6$	  & -			& -         & -    	\\
NGC4517  & ACS/WFC    & F814W   & F606W   & 1.14   & 1.07   & 0.81   & 3.51$\times$10$^6$     & 2.86$\times$$10^6$ & 2.78      & -2.77    \\
NGC4592  & WFC3/UVIS  & F814W   & F555W   & 1.59   & 1.48   & 1.43   & 3.30$\times$10$^6$     & 4.71$\times$$10^6$ & 2.42      & -2.90   \\
NGC4600  & ACS/WFC    & F814W   & F606W   & 1.55   & 1.45   & 1.58   & 6.55$\times$10$^6$     & 1.03$\times$$10^7$ & 3.95      & -1.44   \\
NGC4605  & WFC3/UVIS  & F814W   & F555W   & 2.06   & 1.92   & 5.48   & 4.38$\times$10$^6$     & 2.40$\times$$10^7$ & 2.55      & -1.62   \\
NGC4941  & WFPC2      & F814W   & F606W   & 1.57   & 1.58   & 1.55   & 3.88$\times$10$^7$     & 6.02$\times$$10^7$ & 2.11      & -2.53   \\
NGC5055  & WFPC2      & F814W   & F547M   & 2.52   & 2.54   & 0.56   & 9.26$\times$10$^7$     & 5.12$\times$$10^7$ & 2.77      & -2.60   \\
NGC5068  & WFPC2      & F814W   & F606W   & 0.6    & 0.52   & 1.2    & 1.38$\times$10$^6$     & 1.66$\times$$10^6$ & 3.34      & -1.44   \\
NGC5194  & ACS/WFC    & F814W   & F555W   & 1.33   & 1.3    & 1.2    & 1.34$\times$10$^8$	  & 1.61$\times$$10^8$ & -         & -       \\
NGC5195  & ACS/WFC    & F814W   & F555W   & 1.65   & 1.62   & 5.48   & 1.99$\times$10$^6$     & 1.09$\times$$10^7$ & 3.17      & -1.56   \\
NGC5236  & WFC3/UVIS  & F814W   & F555W   & 1.09   & 1.05   & 4.2    & 3.97$\times$10$^5$     & 1.67$\times$$10^6$ & 3.07      & -0.93   \\
NGC5238  & ACS/WFC    & F814W   & F606W   & 0.3    & 0.35   & 0.5    & 1.45$\times$10$^6$     & 7.21$\times$$10^5$	& 3.01      & -2.48   \\
NGC5457  & ACS/WFC    & F814W   & F555W   & 0.72   & 1.01   & 1.8    & 1.21$\times$10$^7$     & 2.18$\times$$10^7$	& 3.18      & -1.79   \\
NGC6503  & WFC3/UVIS  & F814W   & F555W   & 0.42   & 0.54   & 5.48   & 2.18$\times$10$^6$     & 1.19$\times$$10^7$ & 3.25      & -1.59	 \\
NGC7713  & WFPC2      & F814W   & F606W   & 0.9    & 1.28   & 0.32   & 2.06$\times$10$^6$     & 6.53$\times$$10^5$ & 3.96      & -1.89	  \\
\\
\hline
\end{tabular}
\begin{tablenotes}
{\it NSC parameters}: (1):  Galaxy name as described by Hyperleda Column (2), (3) and (4): Camera and filters in which the galaxy colors were estimated. (5): The color from column (3) \& (4) (Filter2 - Filter1). (6) The $V-I$ color estimated from the color in column (5) using transformations as described in Section~\ref{sec:color_masses}. (7) The $M/L$ in the $I$-band estimated using the relations from \citet{roediger15} and described in Section~\ref{sec:galmass}. The median errors are $\sim$12\% (8) Cluster luminosities estimated using the NSC MGEs from Equation~\ref{eq:lum}, with uncertainties $\sim$12\%. (9) NSC masses estimated using the $M/L$ and luminosities, with median uncertainties $\sim$17\%. Column (10) and (11) are the estimated densities of the NSCs at a median effective radius of the NSCs (5~pc), and a power-law $\gamma$ fit to their slope, with median uncertainties of $\sim$8\% and $\sim$10\% respectively. 
\end{tablenotes}
\egroup
\end{threeparttable}
\end{table}

\clearpage

Some galaxies in our sample have caveats related to their modeling, which require special attention to obtain luminosity models as accurately as possible. We describe here comments on individual galaxies regarding any anomalies that we observe and any special techniques we have employed to fit their surface brightness profiles using \texttt{IMFIT}.
\\\\
\textbf{Circinus galaxy}\\\\
This is a nearby Sb spiral galaxy that is located in the galactic plane (b$\sim$ -4$^\circ$) and highly obscured by dust. Due to this, it requires the magnitudes to be estimated correctly by carefully accounting for the foreground extinction. We take the extinction value from \citet{for12}, where they estimate it by constructing SED of Circinus galaxy using 2MASS J, H, K magnitudes, IRAC, and MIPS measurements and fitting it to a Seyfert 2 model template with various A$_v$ values (see Figure~8 in \citet{for12}). We then use this extinction to estimate the galaxy mass in Section~\ref{sec:data}.
There's also an AGN component in this galaxy \citep{baumgartner13}, which can result in an over-estimation of the color of the NSC. There is also a dynamical estimate of BH in this galaxy \citet{saglia16} around $\sim$~10$^6$~M$_\odot$.
\\\\
\textbf{ESO~274-01}\\\\
This is a late-type edge-on galaxy classified as SAb, with no visible bulge. This galaxy is the closest in our sample ($\sim$2.8~Mpc). It also has a galaxy mass of 3$\times$10$^8$M$_\odot$, which is the lowest mass galaxy in our sample and might harbor a black hole of intermediate-mass. It has a bright foreground star very close to the nucleus. We mask the star in our galaxy model fits.  It contains irregular patches of dust and star formation across its disk. There is a centrally concentrated nucleus that is very bright, and the surrounding galaxy has very low surface brightness. We fit a S\'ersic profile for the NSC and fit an exponential disk profile for the rest of the galaxy. 
\\\\
\textbf{IC~5332}\\\\
This is the only galaxy in our sample for which we had to use WFPC2/WF3 data as the data from the WFPC2/PC was on the edge of the chip and not very useful for our models. The pixel scale is 0.1$''$ for this galaxy, and the PSF is undersampled for the WF3 chip. This affects the center of the galaxy and can affect in measuring the profile of NSC accurately. Using this data, the best-fit model contained a high S\'ersic index (n $>$ 8) for the NSC, which mostly might be due to PSF effects. Thus, we assign a quality 2 to the fits.
\\\\
\textbf{NGC~3115}\\\\
This galaxy is classified to be an S0-edge-on morphological type. It has no visible dust, but while fitting the galaxy, we observed a ring-like residual. So, we fit this galaxy with a S\'ersic for the central NSC, an Edge-on ring for a second component, and another S\'ersic for the third component. This provided a better fit for the galaxy with better residuals. It also has a dynamical BH mass estimate of 8.9$\times$10$^8$M$_\odot$ \citep{saglia16}.     
\\\\
\textbf{NGC~3593}\\\\
This is an SAa type galaxy and contains a lot of dust with visible spiral dust lanes. The color estimates on this galaxy were high (F555W--F814W$\sim$2.5) due to reddening. We apply a dust mask to the galaxy, but since the dust is close to the center, the fits were not as good as expected. The central surface brightness was being overestimated in our fits, and we do not trust the fits for this galaxy and assigned a quality of 2 in our fits. 
\\\\
\textbf{NGC~5194}
\\\\
This is the Whirlpool galaxy or M51a. It is an Sbc galaxy that contains a lot of dusty spiral lanes. It is also classified as a Seyfert I galaxy. Galaxies NGC~5195 and NGC~5236 were similar to this galaxy in terms of dust. We used the IR data to estimate the luminosity models in these galaxies. We used F110W images for NGC~5194 and NGC~5195, as F160W was not available. And, we used F160W images for NGC~5236. Since the native pixel scale of IR images in HST data is 0.13$''$ pixel$^{-1}$, we drizzled the individual .flt images to a smaller pixel scale. Pixel scales of 0.08$''$ were observed to have a good PSF, and the data had a good resolution. We adopted this pixel scale for our final drizzled F110W/F160W images. For masking the dust, we used F814W and F555W images to create a color map and chose a threshold value for the color that was suitable to create a mask. This mask was used on the IR images, and we performed \texttt{IMFIT} modeling on the galaxies. The quality of fits was good for the three galaxies, but NGC~5194 was observed to have a very flat NSC density profile, mostly due to the obscuration of the central regions due to dust. Thus, we assigned a quality of 2 for its model fits. 

\end{document}